\documentclass[journal,twoside]{IEEEtran}
\IEEEoverridecommandlockouts
\usepackage{cite}

\ifCLASSINFOpdf
  \usepackage[pdftex]{graphicx}
  \graphicspath{}
  \DeclareGraphicsExtensions{.pdf,.jpeg,.png}
\else
  \usepackage[dvips]{graphicx}
  \graphicspath{}
  \DeclareGraphicsExtensions{.eps}
\fi
\usepackage[cmex10]{amsmath}
\usepackage{amssymb}
\ifCLASSOPTIONcompsoc
  \usepackage[caption=false,font=normalsize,labelfont=sf,textfont=sf]{subfig}
\else
  \usepackage[caption=false,font=footnotesize]{subfig}
\fi
\usepackage{abbrevs}
\usepackage{tikz}

\usepackage{mathrsfs}
\usepackage{multirow}

\usepackage{amsfonts,bm}

\usepackage[]{algorithmic}

\newcommand{\conv}[2]{\left\langle #1 ; #2 \right\rangle}

\newcommand{\mbmoi}{$\text{\textsc{mbm}}_{01}$\@\xspace}

\newtheorem{theorem}{Theorem}

\begin{document}
\tikzstyle{state}=[shape=circle,draw=blue!90,fill=blue!10,line width=1pt]
%
\title{Poisson multi-Bernoulli mixture conjugate prior for multiple extended target filtering}

\author{Karl~Granstr\"om,~\IEEEmembership{Member,~IEEE},
        Maryam~Fatemi,~\IEEEmembership{Member,~IEEE},
				and~Lennart~Svensson,~\IEEEmembership{Senior Member,~IEEE}
\thanks{K. Granstr\"om and L. Svensson are with the Department of Signals and Systems, Chalmers University of Technology, Gothenburg, Sweden. E-mail: \texttt{firstname.lastsname@chalmers.se}. M. Fatemi is with Zenuity, Gothenburg, Sweden. Email: \texttt{maryam.fatemi@zenuity.com}}%
}

%


\maketitle

\begin{abstract}
This paper presents a Poisson multi-Bernoulli mixture (\pmbm) conjugate prior for multiple extended object filtering. A Poisson point process is used to describe the existence of yet undetected targets, while a multi-Bernoulli mixture describes the distribution of the targets that have been detected. The prediction and update equations are presented for the standard transition density and measurement likelihood. Both the prediction and the update preserve the \pmbm form of the density, and in this sense the \pmbm density is a conjugate prior. However, the unknown data associations lead to an intractably large number of terms in the \pmbm density, and approximations are necessary for tractability. A gamma Gaussian inverse Wishart implementation is presented, along with methods to handle the data association problem. A simulation study shows that the extended target \pmbm filter performs well in comparison to the extended target \dglmb and \lmb filters. An experiment with Lidar data illustrates the benefit of tracking both detected and undetected targets.

\end{abstract}


%
\IEEEpeerreviewmaketitle

%

\section{Introduction}
\label{sec:Intro}

Multiple target tracking (\mtt) is the processing of sets of measurements obtained from multiple sources in order to maintain estimates of the targets' current states\footnote{
In MTT, a multi-target \emph{tracker} produces estimates of target trajectories (state sequences), while a multi-target \emph{filter} produces estimates of the current set of targets. In this paper, we focus on filtering, whereas forming target trajectories is outside the scope of the paper.}. Solving the \mtt problem is complicated by the fact that---in addition to noise, missed detections and clutter---the number of targets is unknown and time-varying. Point target \mtt is defined as tracking targets that give rise to at most one measurement per target at each time step, and extended target \mtt is defined as tracking targets that potentially give rise to more than one measurement at each time step, where the set of measurements are spatially distributed around the extended target.

The focus of this paper is on extended targets. A target may give rise to more than one measurement if the resolution of the sensor, the size of the target, and the distance between target and sensor, are such that multiple resolution cells of the sensor are occupied by a single target. Examples of such scenarios include vehicle tracking using automotive radars, tracking of ships with marine radar stations, and person tracking using laser range sensors. An introduction to extended target tracking and a comprehensive overview of the literature is given in \cite{GranstromBR:2017}.

A common extended target measurement model is the inhomogeneous Poisson Point Process (\ppp), proposed in \cite{gilholm_SPIE_2005_extGroupTrack}. At each time step, a Poisson distributed random number of measurements are generated, distributed around the target. For tracking multiple extended targets, random finite sets (\rfss) can be used to model the problem. \rfss and Finite Set Statistics (\fisst) \cite{mahler_book_2007,Mahler:2014} is a theoretically elegant and appealing approach to the \mtt problem where targets and measurements are modelled as random sets. The \ppp extended target model \cite{gilholm_SPIE_2005_extGroupTrack} has been integrated into several computationally feasible \rfs-based filters, e.g., \cite{mahler_FUSION_2009_extTarg,LundquistGO:2013,GranstromLO:2012,GranstromO:2012a,BeardRGVVS:2016,SwainC:2012}.

In Bayesian statistics, the concepts of \emph{conjugacy} and \emph{conjugate prior}, first introduced by Raiffa and Schlaifer \cite{RaiffaS:1961}, are important. Conjugacy in the context of \mtt means that \emph{``if we start with the proposed conjugate initial prior, then all subsequent predicted and posterior distributions have the same form as the initial prior''} \cite[p. 3460]{VoV:2013}. \mtt conjugate priors are of great interest as they provide families of distributions that are suitable to work with when we seek accurate approximations to the posterior distributions. 

Two different kinds of \mtt conjugate priors can be found in the literature: one based on labelled multi-Bernoulli \rfss, called $\delta$-Generalized Labelled Multi-Bernoulli (\dglmb) \cite{VoV:2013}; and another based on Poisson multi-Bernoulli \rfss, called Poisson  Multi-Bernoulli Mixture (\pmbm) \cite{Williams:2015conjprior}. The \pmbm conjugate prior allows an elegant separation of the set of targets into two disjoint subsets: targets that have been detected, and targets that are unknown, i.e., that have not yet been detected. For the \dglmb multi-object density, conjugacy has been shown for both point targets \cite{VoV:2013} and extended targets \cite{BeardRGVVS:2016}; for the \pmbm multi-object density, conjugacy has only been shown for point targets \cite{Williams:2015conjprior}.

The relation between the two point target conjugate priors are explored in \cite{GarciaFernandezWGS:2018}, where it is shown that the \pmbm density has a more efficient structure than the \dglmb density, with fewer hypotheses. A performance comparison of different implementations of point target filters based on \mtt conjugate priors was presented in \cite{XiaGSGF:2017}, and it showed that the filters based on the \pmbm conjugate prior in general compare well to the filters based on the \dglmb conjugate prior, in terms of tracking performance\footnote{Tracking performance is measured by the localisation error, number of false targets, and number of missed targets, see, e.g.,  \cite[Sec. 13.6]{BlackmanP:1999}.} and computational cost. It is therefore of interest to prove conjugacy for the \pmbm multi-object density also for the standard extended target likelihood \cite{gilholm_SPIE_2005_extGroupTrack}, and to implement a \pmbm filter for extended targets and compare its performance to the \dglmb filter for extended targets.

In this paper, we derive a \pmbm \mtt conjugate prior for the \ppp measurement model \cite{gilholm_SPIE_2005_extGroupTrack} and the standard multi-target motion model, see, e.g., \cite[p. 314]{mahler_book_2007}. A preliminary version of this work was presented in \cite{GranstromFS:2016fusion}. This paper is a significant extension of that work, and contains the following contributions:
\begin{enumerate}
	\item In Section~\ref{sec:conjPrior}, we derive, for the \ppp extended target likelihood, the conjugate update for the \pmbm density, and we review the conjugate prediction for the \pmbm density, which was presented in \cite{Williams:2015conjprior}.
	\item In Section~\ref{sec:AnalysisComplApproxErrorConv}, we analyse the complexity of the \pmbm filter, discuss how the data association problem can be handled, and analyse the approximation error that is incurred by approximating the data association. In Section~\ref{sec:MBmerging}, we propose a merging algorithm that can be used to reduce the number of components in a multi Bernoulli mixture.
	\item In Section~\ref{sec:GGIWPMBM}, we present a computationally feasible implementation of the \pmbm filter, based on gamma Gaussian inverse Wishart (\ggiw) single target models.
	\item In Section~\ref{sec:SimulationStudy}, we present a simulation study, where the \ggiw-\pmbm filter is compared to state-of-the-art algorithms, and we present an experiment, in which the benefits of modelling the targets that have not yet been detected is highlighted.
\end{enumerate}
Problem formulation and modelling are presented in Section~\ref{sec:PaperOverview} and Section~\ref{eq:RFSmodeling}, respectively. The paper is concluded in Section~\ref{sec:Conclusions}.


\section{Problem formulation}
\label{sec:PaperOverview}
The set of targets at time step $k$ is denoted $\setX_{k}$, and is modelled as a \rfs, meaning that the target set cardinality $|\setX_k|$ is a time-varying discrete random variable, and each target state is a random variable. The target state models both kinematic properties (position, velocity, turn-rate, orientation, etc) and target extent (shape and size). The set of measurements at time step $k$ is an \rfs denoted $\setZ_{k}$. There are two types of measurements: clutter measurements and target originated measurements, and the measurement origin is assumed unknown. Further, $\setZ^{k}$ denotes all measurement sets $\setZ_{t}$ from time $t=0$ up to, and including, time $t=k$.

The multi-object posterior density at time $k$, given all measurement sets up to and including time step $k$, is denoted $f_{k|k}(\setX_{k}|\setZ^{k})$. The multitarget Bayes filter propagates in time the multi-target set density $f_{k-1|k-1}(\setX_{k-1} | \setZ^{k-1})$ using the Chapman-Kolmogorov prediction
\begin{subequations}
\begin{align}
& f_{k|k-1}(\setX_{k}|\setZ^{k-1})  \label{eq:MultiobjectPrediction} \\
& = \int f_{k,k-1}(\setX_{k}|\setX_{k-1})f_{k-1|k-1}(\setX_{k-1}|\setZ^{k-1})\delta\setX_{k-1}, \nonumber
\end{align}%
and then updates the density using the Bayes update
\begin{align}
f_{k|k}(\setX_k|\setZ^k) & = \frac{f_{k}(\setZ_k|\setX_k)f_{k|k-1}(\setX_k|\setZ^{k-1})}{\int f_{k}(\setZ_k|\setX_k)f_{k|k-1}(\setX_k|\setZ^{k-1}) \delta\setX_k},
\label{eq:MultiobjectCorrection}
\end{align}
\label{eq:MultiObjectBayesRecursion}%
\end{subequations}
where $f_{k+1,k}(\setX_{k+1}|\setX_{k})$ is the multi-object transition density, $f_{k}(\setZ_k|\setX_k)$ is the multitarget measurement set density, and the integrals are set-integrals, defined in \cite[Sec. 11.3.3]{mahler_book_2007}. In this paper, we model the measurement set density $f_{k}(\setZ_k|\setX_k)$ using the standard \ppp extended target measurement model \cite{gilholm_SPIE_2005_extGroupTrack} and a standard \ppp clutter model. The multi-object transition density $f_{k+1,k}(\setX_{k+1}|\setX_{k})$ is modeled by a standard multi-object Markov density with \ppp birth. 

Among \rfs based filters, there are two main filter types that implement the Bayes recursion \eqref{eq:MultiObjectBayesRecursion} for the multi-object density. The first is based on moment approximations, e.g., the \phd filter and the \cphd filter. The second is based on parameterised density representations, e.g., the \glmb filter and the \pmbm filter. 

The main objectives of this paper are: 1) to show that the \pmbm representation of $f_{k|k}(\setX_{k} | \setZ^{k})$ is an \mtt conjugate prior for the standard extended target tracking models by deriving the corresponding prediction and update, 2) to show how this \pmbm filter can be implemented in a computationally tractable way, and 3) to evaluate the performance of the implementation and compare to state-of-the-art algorithms.


\section{Modeling}
\label{eq:RFSmodeling}
An introduction to \rfss is given in, e.g., \cite{mahler_book_2007}, and an introduction to extended object modelling is given in, e.g., \cite{GranstromBR:2017}. This section first presents a review of random set theory; specifically the \ppp and the Bernoulli process. The standard extended target measurement and motion models are then presented. Notation is given in Table~\ref{tab:notation}.

\begin{table}[t]%
\caption{Notation}%
\label{tab:notation}%
\vspace{-5mm}
\hrulefill
\begin{itemize}
	\item Minor non-bold letters, e.g., $\alpha,\beta,\gamma$, denote scalars, minor bold letters, e.g., $\sx,\sz,\bm{\xi}$, denote vectors, capital non-bold letters, e.g., $\ext,H,F$ denote matrices, and capital bold letters, e.g., $\setX,\setZ,\setC$, denote sets.
	\item $|V|$: determinant of matrix $V$.
	\item $|\setX|$: cardinality of set $\setX$, i.e., number of elements in set $\setX$.
	\item $\mathbf{I}_{m}$: identity matrix of size $m \times m$.
	\item $\conv{a}{b} = \int a(x)b(x)\diff x$: inner product of $a(x)$ and $b(x)$
	\item $h^{\setX} = \prod_{\sx\in\setX} h(\sx)$, where $h^{\emptyset}=1$ by definition.	
	\item $\sum_{\uplus_{i\in\mathbb{I}}\setX^{i}=\setX}$: a sum over all (possibly empty) subsets $\setX^{i}$, $i\in\mathbb{I}$, that are mutually disjoint, and whose union is $\setX$, see \cite[Sec. 11.5.3]{mahler_book_2007}.
	
	\item $\binom{a}{b} = \frac{a!}{b! (a-b)!}$: Binomial coefficient
	\item $\stirlingii{a}{b} = \frac{1}{b!} \sum_{k=0}^{b} (-1)^{b-k} \binom{b}{k} k^{a}$: Stirling number of the second kind
	\item $B(n)$: Bell number of $n$:th order

\end{itemize}%
\vspace{-2mm}
\hrulefill
\end{table}

\subsection{Review of random set modeling}
\subsubsection{Poisson point process}
\label{sec:PPP}
A \ppp is a type of \rfs whose cardinality is Poisson distributed, and all elements (e.g., target states) are independent and identically distributed (iid). A \ppp can be parameterised by an intensity function $D(\sx)$, defined on single target state space. The intensity function can be broken down into two parts $D(\sx)=\mu f(\sx)$: the scalar Poisson rate $\mu > 0$ and the spatial distribution $f(\sx)$. One important property of the intensity is that $\int_{\sx \in S} D(\sx) \diff \sx$ is the expected number of set members in $S$. This can be interpreted to mean that in parts of the state space with high/low intensity $D(\sx)$, there is a high/low chance that set members are located. The \ppp density is
\begin{align}
	f(\setX) = e^{-\conv{D}{1}} \prod_{\sx\in\setX} D(\sx) =  e^{-\mu}\prod_{\sx\in\setX} \mu f(\sx) . \label{eq:PPPdensity}
\end{align}
In this work, {\ppp}s are used to model clutter measurements, extended target measurements, target birth, and undetected targets.

\subsubsection{Bernoulli process}
\label{sec:BernoulliRFS}
A Bernoulli \rfs $\setX$ is a type of \rfs that is empty with probability $1-r$ or, with probability $r$, contains a single element with pdf $f(\sx)$. The cardinality is therefore Bernoulli distributed with parameter $r \in [0,1]$. The Bernoulli density is
\begin{align}
	f(\setX) = \left\{ \begin{array}{ll} 1-r & \setX=\emptyset \\ r  f(\sx) & \setX=\{\sx\} \\ 0 & |\setX|\geq 2 \end{array} \right. \label{eq:BernoulliSetDensity}
\end{align}
In \mtt, a Bernoulli \rfs is a natural representation of a single target, as it captures both the uncertainty regarding the target's existence (via the parameter $r$), as well as the uncertainty regarding the target's state $\sx$ (via the density $f(\sx)$).

For an index set $\mathbb{I}$, a multi Bernoulli (\mb) \rfs $\setX$ is the union of a fixed number of independent Bernoulli {\rfs}s $\setX^i$, $i\in\mathbb{I}$, where $\setX^{i} \cap \setX^{j} = \emptyset$ for all $i,j\in\mathbb{I}$, and $\cup_{i\in\mathbb{I}} \setX^i = \setX$.
The \mb density for a set $\setX$ can be expressed as
\begin{align}%
	f(\setX) = & \left\{\begin{array}{lll} \sum_{ \uplus_{i\in\mathbb{I}}\setX^{i}=\setX} \prod_{i\in\mathbb{I}} f^{i}(\setX^{i}) & \text{if} &  |\setX|\leq |\mathbb{I}|, \\ 0 & \text{if} &  |\setX|>|\mathbb{I}| . \end{array} \right. \label{eq:MultiBernoulliDensity}%
\end{align}%
where the notation $\uplus$ is defined in Table~\ref{tab:notation}. The \mb distribution is defined entirely by the parameters $\{r^{i},f^{i}\}_{i\in\mathbb{I}}$ of the involved Bernoulli {\rfs}s.

Lastly, an \mb mixture (\mbm) density is an \rfs density that is a normalized, weighted sum of \mb densities. In \mtt the weights typically correspond to the probability of different data association sequences. An \mbm is defined entirely by the set of parameters $\{(\mathcal{W}^{j},\{r^{j,i},f^{j,i}\}_{i\in\mathbb{I}^{j}})\}_{j\in\mathbb{J}}$, where $\mathbb{J}$ is an index set for the {\mb}s in the \mbm (also called components of the \mbm), $\mathbb{I}^{j}$ is an index set for the Bernoullis in the $j$th \mb, and $\mathcal{W}^{j}$ is the probability of the $j$th MB. 


\subsection{Standard extended target measurement model}
\label{sec:StandardMeasurementModel}
The set of measurements $\setZ_{k}$ is the union of a set of clutter measurements and a set of target generated measurements; the sets are assumed independent. The clutter is modelled as a \ppp with intensity $\kappa(\sz)=\lambda c(\sz)$. An extended target with state $\sx$ is detected with state dependent probability of detection $p_{\rm D}(\sx)$, and, if it is detected, the target measurements are modelled as a \ppp with intensity $\gamma(\sx)\phi(\sz | \sx)$, where both the Poisson rate $\gamma(\sx)$ and the spatial distribution $\phi(\sz|\sx)$ are state dependent. A \ppp with a probability of detection is sometimes called zero-inflated \ppp.

For a non-empty set of measurements ($|\setZ|>0$), the conditional extended target measurement set likelihood is the product of the probability of detection and the \ppp density,
\begin{align}
	\ell_{\setZ}(\sx) = & p_{\rm D}(\sx) p(\setZ|\sx) =  p_{\rm D}(\sx) e^{-\gamma(\sx)}\prod_{\sz\in\setZ} \gamma(\sx)\phi(\sz|\sx) .\label{eq:CondETTmeasLik}
\end{align}
The effective probability of detection for an extended target with state $\sx$ is $p_{\rm D}(\sx)(1-e^{-\gamma(\sx)})$, where $1-e^{-\gamma(\sx)}$ is the Poisson probability of generating at least one detection.
Accordingly, the effective probability of missed detection, i.e., the probability that the target is not detected, is
\begin{align}
q_{\rm D}(\sx) = 1-p_{\rm D}(\sx) + p_{\rm D}(\sx) e^{-\gamma(\sx)} . \label{eq:EffProbOfMissedDet}
\end{align}
Note that $q_{\rm D}(\sx)$ is the conditional likelihood for an empty set of measurements, i.e., $\ell_{\emptyset}(\sx) = q_{\rm D}(\sx)$ (cf. \eqref{eq:CondETTmeasLik}).

Because of the unknown measurement origin\footnote{An inherent property of \mtt is that it is unknown which measurements are from targets and which are clutter, and among the target generated measurements it is unknown which target generated which measurement. Hence, the update must handle this uncertainty.}, it is necessary to discuss data association. Let the measurements in the set $\setZ$ be indexed by $m\in\mathbb{M}$,
\begin{align}
	\setZ = \left\{\sz^m\right\}_{m\in\mathbb{M}}
\end{align}
and let $\assocspace^{j}$ be the space of all data associations $\assoc$ for the $j$th predicted global hypothesis, i.e., the $j$th predicted \mb. A data association $\assoc\in\assocspace^{j}$ is an assignment of each measurement in $\setZ$ to a source, either to the \emph{background} (clutter or new target) or to one of the existing targets indexed by $\mathbb{I}^{j}$. Note that $\mathbb{M} \cap \mathbb{I}^{j} = \emptyset$ for all $j$.

The space of all data associations for the $j$th hypothesis is $\assocspace^{j} = \altpartition(\mathbb{M} \cup \mathbb{I}^{j})$, i.e., a data association $\assoc\in\assocspace^{j}$ is a partition of $\mathbb{M} \cup \mathbb{I}^{j}$ into non-empty disjoint subsets $C\in\assoc$, called index cells\footnote{For example, let $\mathbb{M}=\left(m_1,m_2,m_3\right)$ and $\mathbb{I}=\left( i_{1} , i_{2} \right)$, i.e., three measurements and two targets. One valid partition of $\mathbb{M}\cap\mathbb{I}$, i.e., one of the possible associations, is $\{m_1,m_2,i_{1}\},\{m_3\},\{i_{2}\}$. The meaning of this is that measurements $m_1,m_2$ are associated to target $i_{1}$, target $i_{2}$ is not detected, and measurement $m_3$ is not associated to any previously detected target, i.e., measurement $m_3$ is either clutter or from a new target.}. Due to the standard \mtt assumption that the targets generate measurements independent of each other, an index cell contains at most one target index, i.e., $| C\cap\mathbb{I}^{j} | \leq 1$ for all $C\in\assoc$. Any association in which there is at least one cell, with at least two target indices, will have zero likelihood because this violates the independence assumption. If  the index cell $C$ contains a target index, then let $i_C$ denote the corresponding target index. Further, let $\altcell_{C}$ denote the measurement cell that corresponds to the index cell $C$, i.e., the set of measurements
\begin{align}
	\altcell_{C} = \bigcup_{m \in C\cap\mathbb{M}} \sz^{m}. \label{eq:MeasurementCell}
\end{align} 


\subsection{Standard dynamic model}
\label{sec:StandardDynamicModel}

The existing targets---both the detected and the undetected---survive from time step $k$ to time step $k+1$ with state dependent probability of survival $p_{\rm S}(\sx_{k})$. The target states evolve independently according to a Markov process with transition density $f_{k+1,k}(\sx_{k+1} | \sx_{k})$. New targets appear independently of the targets that already exist. The target birth is assumed to be a \ppp with intensity $D_{k+1}^{b}(\sx)$.  In this work, target spawning is omitted; for work on spawning in an extended target context see \cite{GranstromO:2013spawn}.



\section{Poisson multi-Bernoulli Mixture filter}
\label{sec:conjPrior}

In this section, the \pmbm conjugate prior for the standard extended object measurement and motion models are presented. Throughout the section time indexing is omitted for the sake of brevity.

\subsection{PMBM density}
\label{sec:PMBMdensity}
The \pmbm model is a combination of a \ppp and a \mbm, where the \ppp describes the distribution of the targets that are thus far undetected, and the \mbm describes the distribution of the targets that have been detected at least once. Thus, the set of targets $\setX$ can be divided into two disjoint subsets, 
\begin{align}
\setX^{u}, \, \setX^{d} : \setX^{u} \cup \setX^{d} = \setX, \, \setX^{u} \cap \setX^{d} = \emptyset , \label{eq:TargetSetDefinition}
\end{align}
corresponding to unknown targets $\setX^{u}$, and detected targets $\setX^{d}$. The \pmbm set density can be expressed as
\begin{subequations}
\begin{align}%
	f(\setX ) & = \sum_{\setX^{u} \uplus \setX^{d} = \setX} f^{u}(\setX^{u} ) \sum_{j\in\mathbb{J}} \mathcal{W}^{j} f^{j}(\setX^d ) ,  \\
	f^{u}(\setX^{u} ) & = e^{-\conv{D^{u}}{1}} \prod_{\sx\in\setX^{u}} D^{u}(\sx) , \\
	f^{j}(\setX^d) & =  \sum_{ \uplus_{i\in\mathbb{I}^{j}} \setX^{i}=\setX^d } \prod_{i\in\mathbb{I}^{j}} f^{j,i} \left( \setX^{i} \right) ,
\end{align}%
\label{eq:PMBMsetDensity}%
\end{subequations}%
where $f^{j,i} \left( \cdot \right)$ are Bernoulli set densities, defined in \eqref{eq:BernoulliSetDensity}. There are $|\mathbb{J}|$ components in the \mb mixture, the $j$th component has $|\mathbb{I}^j|$ Bernoulli components, and the probability of the $j$th \mb component is $\mathcal{W}^j$. In target tracking each of the \mb components in the mixture corresponds to a unique {\em global hypothesis} for the detected targets, i.e., a particular history of data associations for all detected targets. 

The \pmbm density is defined entirely by the parameters 
\begin{align}
D^{u} , \{ (\mathcal{W}^{j}, \{(r^{j,i},f^{j,i})\}_{i\in\mathbb{I}^{j}})\}_{j\in\mathbb{J}}.
\end{align} 
Since the \pmbm density is a \mtt conjugate prior, performing prediction and update means that we compute the new \pmbm density parameters.


\subsection{PMBM filter recursion}
The \pmbm filter consist of a prediction and an update step. The \pmbm conjugate prediction is presented in Theorem~\ref{thm:Prediction}.
\begin{theorem}
\label{thm:Prediction}
\it
Given a posterior \pmbm density with parameters
\begin{align}
	D_{}^{u} , \{ (\mathcal{W}_{}^{j}, \{(r_{}^{j,i},f_{}^{j,i})\}_{i\in\mathbb{I}_{}^{j}})\}_{j\in\mathbb{J}_{}} ,
\end{align} 
and the standard dynamic model (Section~\ref{sec:StandardDynamicModel}), the predicted density is a \pmbm density with parameters
\begin{align}
	D_{+}^{u} , \{ (\mathcal{W}_{+}^{j}, \{(r_{+}^{j,i},f_{+}^{j,i})\}_{i\in\mathbb{I}^{j}})\}_{j\in\mathbb{J}},
\end{align}
where
\begin{subequations}
\begin{align}%
	D_{+}^{u} (\sx) & = D^{b}(\sx) + \conv{D_{}^{u}}{p_{\rm S} f_{k+1,k}} , \label{eq:PredictionUndetected} \\
	 r_{+}^{j,i} & = \conv{ f_{}^{j,i}}{p_{\rm S}} r^{j,i} , \label{eq:PredictedBernoulliProbEx} \\
	 f_{+}^{j,i}(\sx) & = \frac{\conv{ f_{}^{j,i}}{p_{\rm S} f_{k+1,k}}}{\conv{ f_{}^{j,i}}{p_{\rm S}}} , \label{eq:PredictedBernoulliSpatialDistribution}
\end{align}%
\end{subequations}%
and $\mathcal{W}_{+}^{j} = \mathcal{W}_{}^{j} $.
\hfill$\square$
\end{theorem}

The proof of the theorem is omitted for brevity, details can be found in, e.g., \cite{Williams:2015conjprior}.
The \pmbm conjugate update is presented in Theorem~\ref{thm:Update}. 
\begin{theorem}
\label{thm:Update}
\it
Given a prior \pmbm density with parameters
\begin{align}
	D_{+}^{u} , \{ (\mathcal{W}_{+}^{j}, \{(r_{+}^{j,i},f_{+}^{j,i})\}_{i\in\mathbb{I}_{+}^{j}})\}_{j\in\mathbb{J}_{+}},
\end{align}
a set of measurements $\setZ$, and the standard measurement model (Section~\ref{sec:StandardMeasurementModel}), the updated density is a \pmbm density
\begin{subequations}
\begin{align}
	f(\setX|\setZ) & = \sum_{\setX^{u} \uplus \setX^{d} = \setX} f^{u}(\setX^{u}) \sum_{j\in\mathbb{J}_{+}} \sum_{\assoc\in\assocspace^{j}} \mathcal{W}_{\assoc}^{j} 	f_{\assoc}^{j}(\setX^{d}) , \\ 
	f^{u}(\setX^{u}) & = e^{-\conv{D^{u}}{1}} \prod_{\sx\in\setX^{u}} D^{u}(\sx) , \\
	f_{\assoc}^{j}(\setX^{d}) & = \sum_{ \uplus_{C\in\assoc}\setX^{C} = \setX } \prod_{C\in\assoc} f_{C}^{j}(\setX^{C}) ,
\end{align}%
\label{eq:UpdatedPMBMdensity}%
\end{subequations}%
where the weights are
\begin{subequations}
\begin{align}
	& \mathcal{W}_{\assoc}^{j} = \frac{\mathcal{W}_{+}^{j} \prod_{C\in\assoc} \mathcal{L}_C }{ \sum_{j'\in\mathbb{J}} \sum_{\assoc'\in\assocspace^{j'}} \mathcal{W}_{+}^{j'} \prod_{C'\in\assoc'} \mathcal{L}_{C'} } , \label{eq:AssociationWeight}\\
	& \mathcal{L}_{C} = \left\{ \begin{array}{cl} \text{\footnotesize $\kappa^{\altcell_C} + \conv{ D_{+}^{u} }{ \ell_{\altcell_C} }$} & \text{\footnotesize if $C\cap\mathbb{I}^{j} = \emptyset, |\altcell_{C}| = 1,$} \\ \text{\footnotesize $\conv{ D_{+}^{u} }{ \ell_{\altcell_C} }$}  & \text{\footnotesize if $C\cap\mathbb{I}^{j} = \emptyset, |\altcell_{C}| > 1,$} \\  \text{\footnotesize $1-r_{+}^{j,i_{C}}+r_{+}^{j,i_{C}} \conv{ f_{+}^{j,i_{C}} }{ q_{\rm D} }$} & \text{\footnotesize if $C\cap\mathbb{I}^{j}\neq\emptyset, \altcell_{C}=\emptyset,$} \\ \text{\footnotesize $r_{+}^{j,i_{C}} \conv{f_{+}^{j,i_{C}}}{ \ell_{\altcell_C}}$}  & \text{\footnotesize if $C\cap\mathbb{I}^{j}\neq\emptyset, \altcell_{C}\neq\emptyset ,$} \end{array} \right.
\end{align}
the densities $f_{C}^{j}(\setX)$ are Bernoulli densities with parameters
\begin{align}
	& r_{C}^{j} = \left\{ \begin{array}{cl} \frac{ \conv{ D_{+}^{u} }{ \ell_{\altcell_{C}} } }{\kappa^{\altcell_{C}}+\conv{ D_{+}^{u} }{ \ell_{\altcell_{C}} }} & \text{\footnotesize if $C\cap\mathbb{I}^{j} = \emptyset, |\altcell_{C}| = 1$,} \\ 1 &  \text{\footnotesize if $C\cap\mathbb{I}^{j} = \emptyset, |\altcell_{C}| > 1$,} \\  \frac{r_{+}^{j,i_{C}}\conv{ f_{+}^{j,i_{C}} }{ q_{\rm D} }}{1-r_{+}^{j,i_{C}}+r_{+}^{j,i_{C}}\conv{ f_{+}^{j,i_{C}} }{ q_{\rm D} }} & \text{\footnotesize if $C\cap\mathbb{I}^{j}\neq\emptyset, \altcell_{C}=\emptyset$,} \\ 1 & \text{\footnotesize if $C\cap\mathbb{I}^{j}\neq\emptyset, \altcell_{C}\neq\emptyset $,} \end{array} \right. \label{eq:NewBernoulliProbEx}\\
	& f_{C}^{j}(\sx) = \left\{ \begin{array}{cl} \frac{ \ell_{\altcell_{C}}(\sx) D_{+}^{u}(\sx) }{ \conv{ D_{+}^{u} }{ \ell_{\altcell_{C}} } } & \text{\footnotesize if $C\cap\mathbb{I}^{j} = \emptyset$,} \\  \frac{ q_{\rm D}(\sx) f_{+}^{j,i_{C}}(\sx)}{\conv{ f_{+}^{j,i_{C}}}{ q_{\rm D}}} & \text{\footnotesize if $C\cap\mathbb{I}^{j}\neq\emptyset, \altcell_{C}=\emptyset$,} \\ \frac{ \ell_{\altcell_{C}}(\sx) f_{+}^{j,i_{C}} (\sx)}{\conv{ f_{+}^{j,i_{C}} }{ \ell_{\altcell_{C}} }} & \text{\footnotesize if $C\cap\mathbb{I}^{j}\neq\emptyset, \altcell_{C}\neq\emptyset $,} \end{array} \right.  \label{eq:NewBernoulliSpatialDensity}
\end{align}
\label{eq:UpdatedPGFLprevdet}%
\end{subequations}%
and the updated \ppp intensity is $D^{u}(\sx) = q_{\rm D}(\sx)D_{+}^{u}(\sx)$.
\hfill$\square$
\end{theorem}

The proof of the theorem can be found in Appendix~\ref{app:UpdateProof}. By comparing \eqref{eq:UpdatedPMBMdensity} with the \pmbm density \eqref{eq:PMBMsetDensity}, we can immediately identify that we have a \pmbm density. The number of components in the \mbm increases, and contains one \mb for every pair of predicted \mb, $j \in \mathbb{J}_{+}$, and possible association, $\assoc\in\assocspace^{j}$.

\section{Complexity, data association approximation, and approximation error}
\label{sec:AnalysisComplApproxErrorConv}

Due to the unknown number of data associations, the number of components in the \mbm grows rapidly as more data is observed, and it follows that the number of \pmbm parameters increases. In this section we first discuss the complexity of the \pmbm filter. We then discuss methods that can be used to keep the number of \mbm components at a tractable level, and lastly we discuss the approximation error that this reduction incurs.

\subsection{Complexity}
\label{sec:ComplexityAnalysis}

Each \mb component in the \mbm corresponds to a unique global hypothesis, where a global hypothesis was defined in Section~\ref{sec:PMBMdensity} as \textit{a particular history of data associations for all detected targets}. The \pmbm prediction preserves the number of {\mb}s and the number of Bernoullis (see Theorem~\ref{thm:Prediction}), however, due to the unknown data association, the update increases both these numbers. 

In this section, we first give expressions for the number of possible data associations for a predicted \mb, i.e., an expression for the cardinality of the set of data associations $\assocspace^{j}$. Next, we present expressions for the number of \mb components in the updated \pmbm density, and for the number of unique Bernoulli components in the \pmbm. Both the number of updated {\mb}s and the number of unique updated Bernoullis contribute to the computational complexity: in theory, we should compute the probability of each of these {\mb}s, and perform prediction and update operations for every unique Bernoulli. Lastly, we discuss the complexity of the \pmbm filter relative that of the \dglmb filter.

\subsubsection{Number of data associations}
Consider the $j$th predicted \mb with Bernoullis indexed by $\mathbb{I}^{j}$, and a set of detections $\setZ$. 
The number of possible ways in which the $|\setZ|$ detections can be associated to either the $|\mathbb{I}^{j}|$ previously detected objects, or to the background (undetected object or false alarm), i.e., the size of the association space $\assocspace^{j}$, is \cite[Sec. 5]{GranstromSRXF:2018}
\begin{subequations}%
\begin{align}
	N^{\assocspace^j}(|\setZ| , |\mathbb{I}^{j}|) = \sum_{C=1}^{|\setZ|} \stirlingii{|\setZ|}{C} \sum_{T=0}^{\min(C,|\mathbb{I}^{j}|)} \binom{C}{T} \frac{|\mathbb{I}^{j}| !}{(|\mathbb{I}^{j}| -T)!},
	\label{eq:DAcomplexity}
\end{align}
where $\stirlingii{\cdot}{\cdot}$ and $\binom{\cdot}{\cdot}$ are defined in Table~\ref{tab:notation}. The complexity of the update is between exponential $\mathcal{O}(2^{|\setZ|+|\mathbb{I}^{j}|})$ and factorial $\mathcal{O}((|\setZ|+|\mathbb{I}^{j}|)!)$.
For a predicted \pmbm, indexed by $\mathbb{J}$, it follows that the total number of possible associations is
\begin{align}
	N^{\assocspace} = \sum_{j\in\mathbb{J}} N^{\assocspace^j}(|\setZ|,|\mathbb{I}^{j}|).
	\label{eq:total_number_of_associations}
\end{align}%
\label{eq:Number_of_data_associations}%
\end{subequations}

\subsubsection{Number of components in MBM}
Using \eqref{eq:Number_of_data_associations}, we can recursively analyze how the number of global hypotheses changes with time. However, under certain conditions, it is possible to directly obtain an expression for the number of global hypotheses ({\mb}s in the \pmbm) at time $k$.

Let the probabilities of detection and survival be non-zero, $p_{\rm D}(\sx) \in (0,1]$ and $p_{\rm S}(\sx)\in(0, 1]$, respectively. Let the birth intensity $D^{b}(\sx)>0$ and/or the initial undetected intensity $D^{u}(\sx)>0$. This  corresponds to the following: targets can be detected;  an existing target may remain in the surveillance area; new targets may be born; and there may be undetected targets in the surveillance area at initialisation. In summary, this means that at any time step $k$, there may be targets in the surveillance area, that may cause detections.  Lastly, let the \pmbm filter be initialised at time $k=0$ with $\mathbb{J}_{0}=\{j_{1}\}$, $\mathcal{W}_{0}^{j_1}=1$,  and $\mathbb{I}_{0}^{j_{1}}=\emptyset$, i.e., an empty \mbm. This corresponds to zero previously detected targets at initialisation.

Given a measurement set $\setZ_{1}$ at time $k=1$, the number of \mb components in the updated \pmbm density is given by the number of associations,
\begin{align}
	|\mathbb{J}_{1|1}| = N^{\assocspace^{j_1}}(|\setZ_{1}|,0) = \sum_{C=1}^{|\setZ_{1}|} \stirlingii{|\setZ_{1}|}{C} = B(|\setZ_{1}|) ,
\end{align}
where the last equality is a standard relation between the Stiriling numbers and the Bell numbers, see, e.g., \cite{GrahamKP:1988}. In other words, the number of {\mb}s is given by the Bell number of order $|\setZ_{1}|$. It can be shown that the number of \mbm components, given measurement sets up to and including time step $k$ and an empty initial \mbm, is given by the Bell number whose order $n$ is the sum of the measurement set cardinalities,
\begin{align}
	|\mathbb{J}_{k|k}| = |\mathbb{J}_{k+1|k}| = B\left( \sum_{t=1}^{k} \left|\setZ_{t} \right|\right) = B\left(\left|\setZ^{k}\right|\right).
	\label{eq:total_number_of_global_hyps}
\end{align}
Importantly, this is the same as the number of ways that we can partition $\setZ^{k} = \cup_{t=1}^{k}\setZ_{t}$ \cite{FatemiGSRH:2016_PMBradarmapping}. The sequence of Bell numbers $B(n)$ is log-convex\footnote{The sequence of Bell numbers is logarithmically convex, i.e., $B(n)^{2} \leq B(n-1)B(n+1)$ for $n\geq 1$ \cite{Engel:1994}. If the Bell numbers are divided by the factorials, $\frac{B(n)}{n!}$, the sequence is logarithmically concave, $\left(\frac{B(n)}{n!}\right)^{2} \geq \frac{B(n-1)}{(n-1)!} \frac{B(n+1)}{(n+1)!}$, for $n\geq 1$ \cite{Canfield:1995}.}, and $B(n)$ grows very rapidly. For example, for two measurements sets $\setZ_{1}$ and $\setZ_{2}$, both with two measurements, there are $B(2+2)=15$ hypotheses. A small increase in the number of detections per time step to four (twice the amount), results in an \mbm with $B(4+4)=4140$ hypotheses.

Each \mb corresponds to a unique global hypothesis. However, two (or more) {\mb}s may contain identical Bernoulli components, i.e., the histories of data associations may be identical for a pair of Bernoullis in the two {\mb}s. The number of unique Bernoullis at time step $k$ is the number of possible subsets of $\setZ^{k}$:
\begin{align}
	N_{k}^{B} = 2^{\left|\setZ^{k} \right|}.
\end{align}
This describes the number of Bernoulli predictions, cf. \eqref{eq:PredictedBernoulliProbEx} and \eqref{eq:PredictedBernoulliSpatialDistribution}, and Bernoulli updates, cf. \eqref{eq:NewBernoulliProbEx} and \eqref{eq:NewBernoulliSpatialDensity}, that are required in the (exact) \pmbm filter.

\subsubsection{Discussion}

Here we discuss the complexity of the \pmbm filter in relation to that of the \dglmb filter. First, note that the global hypotheses may contain Bernoulli components with uncertain existence, i.e., $r<1$. From each global hypothesis with uncertain target existences, global hypotheses with certain target existence can easily be found. A single Bernoulli with probability of existence $r<1$ and state density $f(\sx)$ can be expanded into a Bernoulli mixture density $f_{\text{ce}}(\cdot)$ that has two hypotheses,
\begin{align}
	f_{\text{ce}}(\setX) = (1-r) f_{1}(\setX) + r f_{2}(\setX),
\end{align}
where $f_{1}(\setX)$ and $f_{2}(\setX)$ are Bernoulli densities with probabilities of existence $r_{1}=0$ and $r_{2}=1$, respectively, and state densities $f_{1}(\sx)=f_{2}(\sx) = f(\sx)$. Generalizing this, an \mb process with $s$ components with uncertain existence (i.e., $r<1$) can be represented by a mixture of $2^s$ \mb processes with certain existences (i.e., each Bernoulli in the \mb has either $r=0$ or $r=1$). In \cite[Sec. 4.A]{GarciaFernandezWGS:2018}, such an \mbm density representation with certain target existence is denoted \mbmoi.

A \dglmb density and a uniquely labelled \mbmoi density can represent the same labelled multi-target densities with the same number of global hypotheses, in which target existence is certain \cite[Prop. 7]{GarciaFernandezWGS:2018}. An \mbmoi has a significantly higher number of hypotheses, compared to the corresponding \mbm \cite[Sec. 4]{GarciaFernandezWGS:2018}; having fewer global hypotheses is advantageous because it translates to a lower computational cost. In the update, the more global hypotheses there are, the more data association weights \eqref{eq:AssociationWeight} have to be computed. Regarding the prediction, the \pmbm prediction can be implemented without approximation; the prediction of the \dglmb density, which has certain target existence, results in an increase of the number of global hypotheses, and thus requires approximation using the $k$-shortest paths algorithm, see \cite{BeardRGVVS:2016} and the discussion in \cite[Sec. 4]{GarciaFernandezWGS:2018}.

For point targets, simulation studies have shown that a better trade-off between tracking performance and computational cost is obtained when global hypotheses with uncertain existence are used \cite{XiaGSGF:2017}. The same conclusion can be drawn for extended targets based on the results of the simulation study presented in Section~\ref{sec:SimulationStudy}.
 
 \subsection{Approximations of the data association problem}
 \label{sec:DAapproximations}
 To achieve computational tractability, it is necessary to reduce the number of \pmbm parameters. Here we will briefly describe the strategy for doing this that was used to obtain the results presented in Section~\ref{sec:SimulationStudy}. First, the number of data associations is reduced using gating, clustering, and ranking of the association events. Second, after an updated \pmbm has been computed, we reduce the number of parameters using pruning, merging, and recycling.
 
 \subsubsection{Reducing the number of associations}
First, gating, described in, e.g., \cite[Sec. 2.2.2.2]{BarShalomB:2000}, is performed; naturally the extended target gates take into account both the position and the extent of the target, as well as state uncertainties. Given the gating, the targets and the measurements are separated into approximately independent sub-groups, using a method similar to the one proposed in \cite[Sec. 3]{ScheelGMRD:2014}. After the grouping, we use the methods proposed in \cite{GranstromLO:2012,GranstromO:2012a} to compute several different partitionings of the measurements. Lastly, for each partitioning we compute the $M$ best assignments using Murty's algorithm  \cite{Murty:1968}. This three step procedure---gating, partitioning, assignment---results in a subset of associations $\hat{\assocspace} \subseteq \assocspace$, and typically reduces the number of associations in the update by several orders of magnitude. Similar approaches to reducing the number of data associations have been used previously in several extended target tracking filters, see \cite{GranstromLO:2012,GranstromO:2012a,LundquistGO:2013,BeardRGVVS:2016}. As an alternative to using partitioning and assignment to find a subset of associations, random sampling methods can be used; this is explored in \cite{GranstromRFS:2017,GranstromSRXF:2018,FatemiGSRH:2016_PMBradarmapping}.

\subsubsection{Reducing the number of parameters}
\label{sec:ReducingPMBMparameters}
After the \pmbm update, \mbm components whose updated weight fall below a threshold are pruned from the \mbm. For the remaining \mbm components, we apply the recycling method suggested in \cite{Williams:2011,Williams:2012}. All Bernoullis with probability of existence below a threshold $\tau_{rec}$ are removed from the \mbm, approximated as a \ppp with intensity $rf(\sx)$, and this intensity is added to the updated undetected \ppp density. If the \ppp intensity is represented by a distribution mixture, which is the typical case, then similar mixture components can be merged, e.g., by minimising the \kldiv, and mixture components with low weights can be pruned from the \ppp intensity. Lastly, we apply the merging algorithm outlined in Section~\ref{sec:MBmerging} to the \mbm.

\subsection{Multi-Bernoulli mixture merging}
\label{sec:MBmerging}

In \cite{Williams:2015} an approximate Poisson Multi-Bernoulli filter for point target tracking is proposed, where the \pmbm density that results after the update is approximated as a \pomb density by using variational approximation to minimise the Kullback-Leibler divergence (\kldiv) between the true \pmbm density and the approximate \pomb density. Empirically, we have found that in extended object filtering it is generally not advisable to merge the whole \pmbm density to a single \pomb density. The main reason is the extent: merging two densities with significantly different extent estimates will result in an approximate density in which the extent estimates are distorted. However, in extended target tracking, similar components in the \pmbm density can be merged, in order to reduce the computational cost of the \pmbm filter.

Consider an \mbm density with \mb components indexed by the index set $\mathbb{J}$. The \kldiv between two multi-Bernoulli densities $j_1\in\mathbb{J}$ and $j_2\in\mathbb{J}$, with equal number of Bernoulli components $|\mathbb{I}^{j_1}| = |\mathbb{I}^{j_2}|$, is upper bounded \cite{Williams:2015}
\begin{align}
	D & \left(f^{j_1} || f^{j_2} \right) \\
	& \leq \sum_{\pi \in \Pi} q(\pi) \prod_{i\in\mathbb{I}^{j_1}}  \int f^{j_1,i}(\setX^{i}) \log \left( q(\pi) \frac{  f^{j_1,i}(\setX^{i})}{f^{j_2,\pi(i)}(\setX^{i}) } \right) \delta \setX^{i} \nonumber
\end{align}
where $\Pi$ is the set of all ways to assign the Bernoulli components indexed by $\mathbb{I}^{j_1}$ to the Bernoulli components indexed by $\mathbb{I}^{j_2}$, and $q(\pi)\in[0,1]$ are weights for the assignments $\pi$, $\sum_{\pi\in\Pi}q(\pi) = 1$; for additional details, see \cite{Williams:2015}.

For two \mb densities, we compute the pairwise \kldiv between the Bernoulli densities, and compute an assignment $\hat{\pi}$ that gives the minimal sum of \kldiv. Setting $q(\hat{\pi})=1$ we get
\begin{align}
	D & \left(f^{j_1} || f^{j_2} \right) \\
	& \leq \prod_{i\in\mathbb{I}^{j_1}}  \int f^{j_1,i}(\setX^{i}) \log \left( \frac{  f^{j_1,i}(\setX^{i})}{f^{j_2,\hat{\pi}(i)}(\setX^{i}) } \right) \delta \setX^{i} \nonumber \\
	& = D_{\rm UB} \left(f^{j_1} || f^{j_2} \right)
\end{align}
where the subscript $\rm UB$ denotes the upper bound. In this work, we use \mbm merging and merge \mb densities for which $D_{\rm UB} \left(f^{j_1} || f^{j_2} \right)$ is smaller than a threshold.
 

\subsection{Approximation error}
\label{sec:ApproxError}
The \pmbm density \eqref{eq:PMBMsetDensity} can be rewritten as a mixture of Poisson Multi-Bernoulli densities,
\begin{align}%
	f(\setX) = & \sum_{j\in\mathbb{J}} \mathcal{W}^{j} \sum_{\setX^{u} \uplus \setX^{d} = \setX} f^{u}(\setX^u) f^{j}(\setX^d),
\end{align}%
where the Poisson density $f^{u}(\setX^u)$ is equal for all components. Using gating, partitioning, and assignment, we seek to prune low weight components from the mixture density, such that only components with significant weights remain. Trivially, pruning updated \mbm components with low weights would achieve precisely this. Let
\begin{align}
	f_{\mathbb{J}}(\setX) = \sum_{j \in \mathbb{J}} \mathcal{W}^{j} f^{j}(\setX) & , & f_{\mathbb{H}}(\setX) = \sum_{j \in \mathbb{H}} \mathcal{W}^{j} f^{j}(\setX) ,
\end{align}
be two unnormalized \pmbm densities with non-negative weights (i.e., the weights do not necessarily sum to one). If $\mathbb{H} \subseteq \mathbb{J}$, then \cite[Prop. 5]{VoVP:2014} shows that the $L_{1}$-error incurred when approximating $f_{\mathbb{J}}(\setX)$ with $f_{\mathbb{H}}(\setX)$ satisfies 
\begin{subequations}
\begin{align}
	\| f_{\mathbb{J}} - f_{\mathbb{H}}  \|_{1} & = \sum_{j\in \mathbb{J}\backslash\mathbb{H}} \mathcal{W}_{j} ,\\
	\left\| \frac{f_{\mathbb{J}}}{\| f_{\mathbb{J}} \|_{1}} - \frac{f_{\mathbb{H}}}{\| f_{\mathbb{H}} \|_{1}}  \right\|_{1} & \leq 2 \frac{\| f_{\mathbb{J}} \|_{1}-\| f_{\mathbb{H}} \|_{1}}{\| f_{\mathbb{J}} \|_{1}} . \label{eq:ApproxErrorUpperBound}
\end{align}
\label{eq:ApproxError}%
\end{subequations}
This result supports the intuitive idea that keeping components with large weights, and discarding components with minimal weights, will yield a small $L_{1}$-error. Further, this shows us that it is possible to achieve an arbitrarily accurate approximation by keeping more components, which in turn shows us that conjugate priors based on \mb densities may be useful even though the theoretical growth of the number of components is hyperexponential. After pruning \pmbm components, the approximate density is normalised. Assuming that $f_{\mathbb{J}}$ is normalised and its approximation $f_{\mathbb{H}}$ is not, \eqref{eq:ApproxErrorUpperBound} shows that the $L_{1}$-error for the normalized approximation is less than two times the sum of the truncated weights. Further analysis of the approximation error is presented in \cite{GranstromSRXF:2018}.

In \cite{Williams:2015conjprior} it is shown that the minimum Kullback-Liebler divergence \ppp approximation of a Bernoulli density is a \ppp whose intensity is equal to the product of the Bernoulli existence probability and state density. In other words, the recycling in Section~\ref{sec:ReducingPMBMparameters} minimises the \kldiv. Setting the recycling threshold $\tau_{rec}=0.1$ is suggested in \cite{Williams:2011,Williams:2012}, where it is shown to give small \kldiv errors. Similarly, by choosing a low threshold in the \mb merging algorithm, we guarantee that the resulting approximation error has low error. Lastly, using a reasoning similar to \eqref{eq:ApproxError}, it can be shown that pruning the \ppp intensity by removing low weight components, and merging similar components by minimising the \kldiv, incurs a small error.


\section{\ggiw implementation}
\label{sec:GGIWPMBM}

In this section, an implementation of the \pmbm filter is presented. There are several single extended target models available in the literature, see \cite{GranstromBR:2017} for an overview. Here we have chosen the random matrix model \cite{Koch:2008,FeldmannFK:2011}, in which the target shape is approximated by an ellipse.  The random matrix model is relatively simple to use, yet flexible enough to be applicable to data from radar \cite{GranstromNBLS:2014,GranstromNBLS:2015_TGARS}, lidar \cite{GranstromO:2012a,WienekeK:2012,GranstromRFS:2017,GranstromSRXF:2018}, and camera \cite{DaveyWV:2013}. With the random matrix model, it is possible to handle noisy non-linear measurement models, e.g., noisy polar measurements \cite{VivoneBGW:2016,VivoneBGNC:2015_ConvMeas,VivoneBGNC:2017,SchusterRW:2015}; in these cases the data is pre-processed with a polar-to-Cartesian transformation. Furthermore, the random matrix model has been used in many other multiple extended target tracking filters, making comparison easy. A comprehensive discussion of the random matrix model is given in \cite[Sec. 3]{GranstromBR:2017}

\subsection{Single target models}
In the random matrix model, the extended target state $\sx_k$ is the combination of the scalar $\gamma_k$, the vector $\bm{\xi}_k$ and the matrix $\ext_k$. The random vector $\bm{\xi}_k \in \mathbb{R}^{n_x}$ is the kinematic state, which describes the target's position and its motion parameters (e.g., velocity, acceleration and turn-rate). The random matrix $\ext_k \in\mathbb{S}_{++}^{d}$ is the extent state and describes the target's size and shape, and $d$ is the dimension of the extent (typically $d=2$ or $d=3$). Lastly, the random variable $\gamma_k > 0$ is the measurement model Poisson rate.

The measurement likelihood for a single measurement $\sz$, cf. \eqref{eq:CondETTmeasLik}, is
\begin{align}
	\phi(\sz_k|\sx_k) &  = \Npdfbig{\sz_k}{H_k\bm{\xi}_k}{\ext_k}, \label{eq:GaussianMeasurementLikelihood}
\end{align}
where $H_k$ is a known measurement model. The single-target conjugate prior for the \ppp model \eqref{eq:CondETTmeasLik} with single measurement likelihood \eqref{eq:GaussianMeasurementLikelihood} is a \ggiw distribution \cite{FeldmannFK:2011,GranstromO:2012c},
\begin{align}
	f_{k|k}\left(\sx\right) =& \Gammapdf{\gamma_k}{\alpha_{k|k}}{\beta_{k|k}} \Npdfbig{\bm{\xi}_k}{\GaussMean_{k|k}}{P_{k|k}} \nonumber \\
	& \times \IWishpdf{\ext_k}{v_{k|k}}{V_{k|k}}, \\
	= & \GGIWpdf{\sx_k}{\zeta_{k|k}},
\end{align}%
where $\zeta_{k|k} = \left\{\alpha_{k|k},\beta_{k|k},m_{k|k},P_{k|k},v_{k|k},V_{k|k}\right\}$ is the set of \ggiw density parameters. The gamma distribution is the conjugate prior for the unknown Poisson rate, and the Gaussian-inverse Wishart distributions are the conjugate priors for Gaussian distributed detections with unknown mean and covariance. 

For a \ggiw distribution with prior parameters $\zeta_{k|k-1}$, that is updated with a set of detections $\setW$ under the linear Gaussian model \eqref{eq:GaussianMeasurementLikelihood}, the updated parameters $\zeta_{k|k}$, and the corresponding predicted likelihood,  are given in Table~\ref{tab:GGIWupdate}. For further discussions about the measurement update within the random matrix extended target model see, e.g., \cite{Koch:2008,FeldmannFK:2011,Orguner:2012}.

\begin{table}
\caption{\ggiw update}
\label{tab:GGIWupdate}
\vspace{-5mm}
\rule[0pt]{\columnwidth}{1pt}
\textbf{Input:} \ggiw parameter $\zeta_{+}$, set of detections $\setW$, measurement model $H$.

\textbf{Output:} Updated \ggiw parameter $\zeta$ and predicted likelihood $\ell$:
\begin{align*}
\zeta = \left\{ \begin{array}{rcl}
\alpha &= &  \alpha_{+} + |\setW|, \\
\beta &= & \beta_{+} + 1,\\
\GaussMean &= & \GaussMean_{+} + K\varepsilon, \\
P &= & P_{+}-K H P_{+},\\
v &= & v_{+} + |\setW|,\\
V &= & V_{+} + N + Z
\end{array}\right.
\end{align*}
where
\begin{align*}
\begin{array}{rcl}
\bar{\sz}  &= &  \frac{1}{|\setW|}\sum_{\sz^{i} \in \setW}{\sz^{i}}, \\
Z &= & \sum_{\sz^{i}\in \setW} \left(\sz^{i} - \bar{\sz}\right)\left(\sz^{i} - \bar{\sz}\right)^{\tp} \\
\hat{\ext} &= & V_{+} \left(v_{+}-2d-2\right)^{-1},\\
\varepsilon &= &  \bar{\sz} - H \GaussMean_{+},\\
S  &= & H P_{+} H^{\tp} + \frac{\hat{\ext}}{|\setW|},\\
K  &= & P_{+}H^{\tp}\left(S\right)^{-1},\\
N &= &  \hat{\ext}^{1/2} S^{-1/2} \varepsilon \varepsilon^{\tp} {S}^{-\tp/2} \hat{\ext}^{\tp/2}
\end{array}
\end{align*}
Predicted likelihood, where $\Gamma(\cdot)$ is the Gamma function, and $\Gamma_{d}(\cdot)$ is the multivariate Gamma function,
\begin{align*}
\ell = \left(\pi^{|\setW|}|\setW|\right)^{-\frac{d}{2}} \frac{\left|V_{+} \right|^{\frac{v_{+}-d-1}{2}} \Gamma_{d}\left(\frac{v-d-1}{2}\right) \left| \hat{\ext} \right|^{\frac{1}{2}} \Gamma\left(\alpha\right)\left(\beta_{+}\right)^{\alpha_{+}}} {\left|V \right|^{\frac{v-d-1}{2}} \Gamma_{d}\left(\frac{v_{+}-d-1}{2}\right) \left| S \right|^{\frac{1}{2}} \Gamma\left(\alpha_{+}\right)\left(\beta\right)^{\alpha} }
\end{align*}
\rule[0pt]{\columnwidth}{1pt}
\end{table}

\begin{table}
\caption{\ggiw prediction}
\label{tab:GGIWprediction}
\vspace{-5mm}
\rule[0pt]{\columnwidth}{1pt}
\textbf{Input:} \ggiw parameter $\zeta$, motion model $g(\cdot)$, process noise covariance $Q$, transformation matrix $M(\cdot)$, sampling time $T_s$, maneuvering correlation constant $\tau$, measurement rate parameter $\eta$.

\textbf{Output:} Predicted \ggiw parameter $\zeta_{+}$, where $G = \left. \nabla_{\bm{\xi}} g(\bm{\xi})\right|_{\bm{\xi}=\GaussMean}$,
\begin{align*}
\zeta_{+} = \left\{\begin{array}{rcl}
\alpha_{+} &= & \frac{\alpha}{\eta},\\
\beta_{+} &= &  \frac{\beta}{\eta},\\
\GaussMean_{+} &= & g\left(\GaussMean\right),\\
P_{+} &= & G P G^{\tp} + Q,\\
v_{+} &= & 2d+2 + e^{-T_s / \tau}\left(v -2d -2\right), \\
V_{+} &= & e^{-T_s / \tau} M\left(\GaussMean\right) V M\left(\GaussMean\right)^{\tp}
\end{array}\right.
\end{align*}
\rule[0pt]{\columnwidth}{1pt}
\end{table}

The motion models are
\begin{subequations}
\begin{align}
\bm{\xi}_{k+1} & = g\left(\bm{\xi}_{k}\right) + \sw_{k}, \\
\ext_{k+1} & = M(\bm{\xi}_{k})\ext_{k}M(\bm{\xi}_{k})^{\tp}, \\
\gamma_{k+1} & = \gamma_{k}.
\end{align}
\end{subequations}
where $g(\cdot)$ is a kinematic motion model, $\sw_{k}$ is Gaussian process noise with zero mean and covariance $Q$, and $M(\bm{\xi}_{k})$ is a transformation matrix. For these motion models, the predicted parameters $\zeta_{k+1|k}$ for a \ggiw distribution with posterior parameters $\zeta_{k|k}$ are given in Table~\ref{tab:GGIWprediction}. For longer discussions about prediction within the random matrix extended target model, see, \egp, \cite{Koch:2008,FeldmannFK:2011,GranstromO:2014}.

\begin{table}
\caption{Assumptions}%
\label{tab:LinGaussAssumptions}%
\vspace{-5mm}
\rule[0pt]{\columnwidth}{1pt}
\begin{itemize}
	\item \ggiw birth \ppp intensity with known parameters,
		\begin{align}
			\begin{array}{c}
				D_{k+1}^{b}(\sx)  = \sum_{n=1}^{N_{k+1}^{b}} w_{k+1}^{b,n}\GGIWpdf{\sx_{k+1}}{\zeta_{k+1}^{b,n}} .
			\end{array}
		t\end{align}
	\item \ggiw initial undetected \ppp intensity with known parameters,
		\begin{align}
			D_{0}^{u}(\sx) & = \sum_{n=1}^{N_{0}^{u}} w_{0}^{u,n}\GGIWpdf{\sx_{0}}{\zeta_{0}^{u,n}} .
		\end{align}
	\item Empty initial \mbm: $\mathbb{J}_{0}=\{j_{1}\}$, $\mathcal{W}_{0}^{j_1}=1$, and $\mathbb{I}_{0}^{j_{1}}=\emptyset$.
	\item Probabilitites of detection and survival can be approximated as
\begin{align}
p_{\rm D}(\sx) f_{}^{}(\sx) \approx &  p_{\rm D}(\hat{\sx}) f_{}^{}(\sx), & p_{\rm S}(\sx) f_{}^{}(\sx) \approx & p_{\rm S}(\hat{\sx}) f_{}^{}(\sx) .
\end{align}
where $\hat{\sx} = \mathbb{E}[\sx] = \int \sx f_{}^{}(\sx)  \diff \sx$. 
	\item Clutter Poisson rate $\lambda$ is known and the spatial distribution is uniform, $c(\sz) = {\rm A}^{-1}$,
		where $\rm A$ is the volume of the surveillance region.
	
\end{itemize}
\vspace{-1mm}
\rule[0pt]{\columnwidth}{1pt}
\end{table}

\subsection{Pseudo code for the update and the prediction}

The \ggiw-\pmbm filter propagates in time the \ggiw-\pmbm density parameters, using a recursion that consists of an update and a prediction. The assumptions are listed in Table~\ref{tab:LinGaussAssumptions}. The assumptions about the probabilities of detection and survival hold trivially if $p_{\rm D}(\cdot)$ and $p_{\rm S}(\cdot)$ are constants, and the assumptions are expected to hold when $p_{\rm D}(\cdot)$ and $p_{\rm S}(\cdot)$ are sufficiently smooth functions within the uncertainty area of the estimate. Note that the assumptions of \ggiw mixture intensities for the birth \ppp and the initial undetected \ppp result in all single target densities in the \pmbm filter being \ggiw densities, due to the conjugacy property.

\begin{table}
	\caption{\ggiw \pmbm prediction}
	\label{tab:GGIWPMBMprediction}
	\vspace{-5mm}
\rule[0pt]{\columnwidth}{1pt}
\textbf{Input:} $D_{}^{u} , \{ (\mathcal{W}_{}^{j}, \{(r_{}^{j,i},f_{}^{j,i})\}_{i\in\mathbb{I}_{}^{j}})\}_{j\in\mathbb{J}_{}}$.

\textbf{Output:}  $D_{+}^{u} , \{ (\mathcal{W}_{+}^{j}, \{(r_{+}^{j,i},f_{+}^{j,i})\}_{i\in\mathbb{I}^{j}})\}_{j\in\mathbb{J}}$

	\begin{align*}%
D_{+}^{u} (\sx_{}) = & \sum_{n=1}^{N^{b}} w^{b,n}\GGIWpdf{\sx_{}}{\zeta^{b,n}} \nonumber\\
& + \sum_{n=1}^{N_{}^{u}} w_{}^{u,n} p_{\rm S}\left(\hat{\sx}_{}^{u,n}\right)\GGIWpdf{\sx}{\zeta_{+}^{u,n}} \\
 r_{+}^{j,i} = & p_{\rm S}\left(\hat{\sx}^{j,i}\right)r_{}^{j,i} \\
 f_{+}^{j,i}(\sx) = & \GGIWpdf{\sx}{\zeta_{+}^{j,i}}
\end{align*}%
and $\mathcal{W}_{+}^{j}=\mathcal{W}_{}^{j}$, where $\zeta_{+}^{u,n}$ and $\zeta_{+}^{j,i}$ are computed as in Table~\ref{tab:GGIWprediction}.

\rule[0pt]{\columnwidth}{1pt}
\end{table}

\begin{table}                      
\caption{\ggiw-\pmbm update}          
\label{tab:GGIWPMBMupdate}                           
\vspace{-5mm}
\rule[0pt]{\columnwidth}{1pt}
\begin{algorithmic}                    
    \STATE \textbf{Input:} Predicted parameters $D_{+}^{u} , \{ (\mathcal{W}_{+}^{j}, \{(r_{+}^{j,i},f_{+}^{j,i})\}_{i\in\mathbb{I}_{+}^{j}})\}_{j\in\mathbb{J}_{+}}$, measurement set $\setZ$
    \STATE \textbf{Output:} Updated parameters $D^{u} , \{ (\mathcal{W}^{j}, \{(r^{j,i},f^{j,i})\}_{i\in\mathbb{I}^{j}})\}_{j\in\mathbb{J}}.$
    \STATE{Compute $D^{u}$ as in Table~\ref{tab:UndetUpdate}}
    \STATE{Initialise: $\mathbb{J}\leftarrow\emptyset$, $j\leftarrow0$}
    \FOR{$j_{+}\in\mathbb{J}_{+}$}
    	\STATE{Compute subset of associations $\hat{\assocspace}^{j_{+}}$}
    	\FOR{$\assoc\in\hat{\assocspace}^{j_{+}}$}
		\STATE{Increment: $j \leftarrow j+1$, $\mathbb{J} \leftarrow \mathbb{J} \cup j$}
		\STATE{Initialise: $\mathbb{I}^{j} \leftarrow \emptyset$, $i\leftarrow0$, $\mathbb{D}\leftarrow\emptyset$, $\mathcal{L}_{\assoc}^{j_{+}}\leftarrow1$}
		\FOR{$C\in\assoc$}
			\STATE{Increment: $i \leftarrow i+1$, $\mathbb{I}^{j} \leftarrow \mathbb{I}^{j} \cup i$}
			\IF{$C\cap\mathbb{I}^{j_{+}}=\emptyset$}
				\STATE{From $r^{j_{+},i_{C}},f^{j_{+},i_{C}}$, compute $r,f,\mathcal{L}$ as in Table~\ref{tab:NewBernoulliUpdate}}
			\ELSE
				\STATE{From $r^{j_{+},i_{C}},f^{j_{+},i_{C}}$, compute $r,f,\mathcal{L}$ as in Table~\ref{tab:BernoulliUpdateDetected}}
				\STATE{$\mathbb{D} \leftarrow \mathbb{D}\cup i_{C}$}
			\ENDIF
			\STATE{$r^{j,i} \leftarrow r$, $f^{j,i} \leftarrow f$, $\mathcal{L}_{\assoc}^{j_{+}} \leftarrow \mathcal{L}_{\assoc}^{j_{+}} \times \mathcal{L}$}
		\ENDFOR
		\FOR{$i_{+} \in (\mathbb{I}^{j_{+}}\backslash\mathbb{D})$}
			\STATE{Increment: $i \leftarrow i+1$, $\mathbb{I}^{j} \leftarrow \mathbb{I}^{j} \cup i$}
			\STATE{From $r^{j_{+},i_{C}},f^{j_{+},i_{C}}$, compute $r,f,\mathcal{L}$ as in Table~\ref{tab:BernoulliUpdateMissed}}
			\STATE{$r^{j,i} \leftarrow r$, $f^{j,i} \leftarrow f$, $\mathcal{L}_{\assoc}^{j_{+}} \leftarrow \mathcal{L}_{\assoc}^{j_{+}} \times \mathcal{L}$}
		\ENDFOR
		\STATE{$\mathcal{W}^{j} \leftarrow \mathcal{L}_{\assoc}^{j_{+}}$}
	\ENDFOR
    \ENDFOR
    \STATE{$\mathcal{W}^{j} \leftarrow \frac{\mathcal{W}^{j} }{\sum_{j'\in\mathbb{J}}\mathcal{W}^{j'} }$}
\end{algorithmic}
\rule[0pt]{\columnwidth}{1pt}
\end{table}

The predicted \ggiw-\pmbm parameters are presented in Table~\ref{tab:GGIWPMBMprediction}. The updated density for the undetected \ppp has $N_{k+1}^{b}+N_{k|k}^{u}$ \ggiw components after the prediction, whereas the  number of \mbm parameters remains the same as before the prediction. The pseudo-code for the \pmbm update, under assumed \ggiw models, is given in Table~\ref{tab:GGIWPMBMupdate}. This builds upon the \ppp intensity updates for missed detection and detection, see Tables~\ref{tab:UndetUpdate} and \ref{tab:NewBernoulliUpdate}, respectively, and the Bernoulli updates for detection and missed detection, see Tables~\ref{tab:BernoulliUpdateDetected} and \ref{tab:BernoulliUpdateMissed}, respectively. 

The density for a target detected for the first time, see Table~\ref{tab:UndetUpdate}, is multimodal, with one mode for each of the \ggiw components in the predicted \ppp intensity $D^{u}$. Mixture reduction can be used to reduce this to a uni-modal \ggiw density \cite{GranstromO:2012e,GranstromO:2012c}. This reduction typically has low error, because one of the modes in the predicted \ppp intensity is typically much likelier than the other modes.

The density for a previously detected target that is now missed, see Table~\ref{tab:BernoulliUpdateMissed}, is multi-modal with two modes. This is due to the fact that there are two ways for the target detection to result in an empty measurement set. The first corresponds to the detection process modeled by $p_{\rm D}(\cdot)$, which may result in a missed detection. The second corresponds to the Poisson number of detections governed by the parameter $\gamma$, i.e., the Poisson random number of detections is zero. Note that the Gaussian and inverse Wishart parameters are identical in both cases, it is only the gamma parameters that differ. Using gamma mixture reduction \cite{GranstromO:2012c}, the bi-modality of the $\gamma_k$ estimate can be reduced to a single mode.

For a predicted global hypothesis and a data association, any cluster of measurements not associated to a predicted target will initiate a new Bernoulli in the updated global hypothesis. After the update, as mentioned in Section~\ref{sec:ReducingPMBMparameters}, we check each updated global hypothesis and any Bernoullis with probability of existence below a threshold $\tau_{rec}$ are pruned.

\begin{table}
	\caption{\ggiw \ppp update: missed detection}
	\label{tab:UndetUpdate}
	\vspace{-5mm}
\rule[0pt]{\columnwidth}{1pt}
\textbf{Input:} Predicted \ppp intensity $D_{+}^{u} (\sx)$, probability of missed detection $q_{\rm D}(\sx)$.

\textbf{Output:} Updated \ppp intensity $D^{u} (\sx)$:

	\begin{align*}
	\begin{array}{c}
			D^{u} (\sx) =  \sum_{n=1}^{N^{u}} \left( w_{1}^{u,n} \GGIWpdf{\sx}{\zeta_{1}^{u,n}} + w_{2}^{u,n} \GGIWpdf{\sx}{\zeta_{2}^{u,n}}\right)
	\end{array}
	\end{align*}
	where 
	\begin{align*}
		w_{1}^{u,n} & =  \left(1 - p_{\rm D}\left(\hat{\sx}^{u,n}\right) \right) w^{u,n} \\
		w_{2}^{u,n} & = p_{\rm D}\left(\hat{\sx}^{u,n}\right) \left(\frac{\beta_{+}^{u,n}}{\beta_{+}^{u,n} + 1}\right)^{\alpha_{+}^{u,n}}  w^{u,n} \\
		\zeta_{1}^{u,n} & = \zeta_{+}^{u,n} \\
		\zeta_{2}^{u,n} & = \{ \alpha_{+}^{u,n} , \beta_{+}^{u,n}+1 , \GaussMean_{+}^{u,n} , P_{+}^{u,n} , v_{+}^{u,n} , V_{+}^{u,n}\}
	\end{align*}
\rule[0pt]{\columnwidth}{1pt}
\end{table}

\begin{table}
	\caption{\ggiw \ppp update: detection}
	\label{tab:NewBernoulliUpdate}
	\vspace{-5mm}
\rule[0pt]{\columnwidth}{1pt}
\textbf{Input:} Predicted \ppp intensity $D_{+}^{u} (\sx)$, and measurements set $\altcell$.

\textbf{Output:} Bernoulli parameters $r_{\altcell}^u$, $f_{\altcell}^u(\sx)$ and likelihood $\mathcal{L}_{\altcell}^u$:

	\begin{align*}
		r_{\altcell}^{u} & =  \left\{ \begin{array}{ccc} 1 & \text{if} & |\altcell| > 1 \\ \frac{\mathcal{L}_{\altcell}}{\kappa^{\altcell}+\mathcal{L}_{\altcell}} & \text{if} & |\altcell| = 1 \end{array} \right. \\
			f_{\altcell}^{u}(\sx) & = \frac{ \sum_{n=1}^{N^{u}} w^{u,n} p_{\rm D}\left(\hat{\sx}^{u,n}\right) \ell_{\altcell}^{u,n}  \GGIWpdf{\sx}{\zeta_{\altcell}^{u,n}}}{ \sum_{n=1}^{N^{u}} w^{u,n} p_{\rm D}\left(\hat{\sx}^{u,n}\right) \ell_{\altcell}^{u,n}  } \\
	\mathcal{L}_{\altcell}^u & = \sum_{n=1}^{N^{u}} w^{u,n} p_{\rm D}\left(\hat{\sx}^{u,n}\right) \ell_{\altcell}^{u,n}
	\end{align*}
	where $\zeta_{\altcell}^{u,n}$ and $ \ell_{\altcell}^{u,n} $ are computed as in Table~\ref{tab:GGIWupdate}. 
	
\rule[0pt]{\columnwidth}{1pt}
\end{table}


\section{Results}
\label{sec:SimulationStudy}

In this section the results from a Monte Carlo simulation study, and experiments with laser range data, are presented.\footnote{All compared filters were implemented in \matlab, and the simulations and experiments were run on a laptop with a $3.1$ GHz Intel Core i7 processor and $16$ Gb memory.} A comparison between the \phd, \cphd, \lmb and \dglmb filters is presented in \cite{BeardRGVVS:2016}. It shows that the \lmb filter and the \dglmb filter outperform the \phd filter and the \cphd filter. Therefore, in the simulation study presented here, we focus on comparing the \pmbm filter to the \dglmb filter and the \lmb filter. In \cite{BeardRGVVS:2016} two variants of the \lmb filter are presented, one with known \mb birth and one with an adaptive birth process. We found that the \lmb filter with adaptive birth process performed better in the simulated scenarios, and have therefore chosen to only present those results. In both the \pmbm filter and the \dglmb filter, the birth processes were tuned to fit the simulated scenarios well. Specifically, the birth processes were represented by a single Gaussian located at the center of the surveillance space, with a covariance chosen such that most of the surveillance area is within three standard deviations. In scenarios where the birth process cannot be tuned to the problem, birth processes with uniform position density can be used in both filters; for details on how to achieve this, see \cite{BeardVVA:2012}. For further details about the \dglmb filter and the \lmb filter, refer to \cite{BeardRGVVS:2016}. 

\begin{table}
	\caption{\ggiw Bernoulli update: detection}
	\label{tab:BernoulliUpdateDetected}
	\vspace{-5mm}
\rule[0pt]{\columnwidth}{1pt}
\textbf{Input:} Predicted Bernoulli parameters $r_{+}^{j,i}$, $f_{+}^{j,i}(\sx)$, and measurement set $\altcell$.

\textbf{Output:} Updated Bernoulli parameters $r_{\altcell}^{j,i} $, $f_{\altcell}^{j,i}(\sx)$, and likelihood $\mathcal{L}_{\altcell}^{j,i}$:

\begin{align*}
	r_{\altcell}^{j,i} & = 1  \\
	f_{\altcell}^{j,i}(\sx) & = \GGIWpdf{\sx}{\zeta_{\altcell}^{j,i}}\\
	\mathcal{L}_{\altcell}^{j,i}  & = r_{+}^{j,i} p_{\rm D}\left(\hat{\sx}^{j,i}\right) \ell_{\altcell}^{j,i}
\end{align*}
where $\zeta_{\altcell}^{j,i}$ and $\ell_{\altcell}^{j,i}$ are computed as in Table~\ref{tab:GGIWupdate}.

\rule[0pt]{\columnwidth}{1pt}
\end{table}

\begin{table}
	\caption{\ggiw Bernoulli update: missed detection}
	\label{tab:BernoulliUpdateMissed}
	\vspace{-5mm}
\rule[0pt]{\columnwidth}{1pt}
\textbf{Input:} Predicted Bernoulli parameters $r^{j,i}$, $f^{j,i}(\sx)$.

\textbf{Output:} Updated Bernoulli parameters $r_{\emptyset}^{j,i} $, $f_{\emptyset}^{j,i}(\sx)$, and likelihood $\mathcal{L}_{\emptyset}^{j,i}$:
 
\begin{align*}
	r_{\emptyset}^{j,i} & = \frac{r^{j,i}q_{\rm D}^{j,i} }{1-r^{j,i}+r^{j,i}q_{\rm D}^{j,i} }  \\
	f_{\emptyset}^{j,i}(\sx) & = w_{1}^{j,i} \GGIWpdf{\sx_k}{\zeta_{1}^{j,i}} + w_{2}^{j,i} \GGIWpdf{\sx_k}{\zeta_{2}^{j,i}}  \nonumber \\
	\mathcal{L}_{\emptyset}^{j,i}  & = 1-r^{j,i}+r^{j,i}q_{\rm D}^{j,i} 
\end{align*}
where 
\begin{align*}
	q_{\rm D}^{j,i}  & = 1 - p_{\rm D}\left(\hat{\sx}^{j,i}\right) + p_{\rm D}\left(\hat{\sx}^{j,i}\right) \left(\frac{\beta^{j,i}}{\beta^{j,i} + 1}\right)^{\alpha^{j,i}} \\
	w_{1}^{j,i} & = \left(q_{\rm D}^{j,i}\right)^{-1}  \left(1 - p_{\rm D}\left(\hat{\sx}^{j,i}\right) \right)  \\
	w_{2}^{j,i} & = \left(q_{\rm D}^{j,i}\right)^{-1}  p_{\rm D}\left(\hat{\sx}^{j,i}\right) \left(\frac{\beta^{j,i}}{\beta^{j,i} + 1}\right)^{\alpha^{j,i}}   \\
	\zeta_{1}^{j,i} & = \zeta^{j,i} \\
	\zeta_{2}^{j,i} & = \{ \alpha^{j,i} , \beta^{j,i}+1 , m^{j,i} , P^{j,i} , v^{j,i} , V^{j,i}\} 
\end{align*}
\rule[0pt]{\columnwidth}{1pt}
\end{table}

The kinematic state is $\bm{\xi}_k=\left[\mathbf{p}_{k},\ \mathbf{v}_{k}\right]^{\tp}\in\mathbb{R}^{4}$ and describes the target's position $\mathbf{p}_{k}\in\mathbb{R}^{2}$ and velocity $\mathbf{v}_{k}\in\mathbb{R}^{2}$. The random matrix $\ext_k\in\mathbb{S}_{++}^{2}$ is two-dimensional. The motion model $g(\cdot)$ and process noise covariance $Q$ are
	\begin{align*}
		& g(\bm{\xi}_{k}) = \begin{bmatrix}
			\mathbf{I}_{2} & T_s \mathbf{I}_{2} \\
			\mathbf{0}_{2} & \mathbf{I}_{2}
		\end{bmatrix}\bm{\xi}_{k}, \ Q = \mathbf{G}\sigma_{a}^{2}\mathbf{I}_{2}  \mathbf{G}^{\tp}, \
		\mathbf{G} = \begin{bmatrix}
			\frac{T_s^2}{2}\mathbf{I}_{2} \\
			T_s\mathbf{I}_{2}
		\end{bmatrix},
		\label{eq:motion_model}%
	\end{align*}%
where $T_s$ is the sampling time and $\sigma_{a}$ is the acceleration standard deviation. Because the kinematic state motion model is constant velocity, the extent transformation function $M$ is an identity matrix, $M(\bm{\xi}_{k})  = \mathbf{I}_2$.

For \ggiw-\pmbm, we use Estimator 1 from \cite[Sec. 6.A]{GarciaFernandezWGS:2018}: an estimate of the set of targets is obtained by taking the mean vector of all Bernoulli estimates with existence probability larger than $0.5$ from the \mb component with largest \mb weight. For \ggiw-\dglmb and \ggiw-\lmb, target extraction is performed analogously, see \cite[Sec. 4.A.3]{BeardRGVVS:2016} or \cite[Sec. 6.E]{VoV:2013} for details. Further discussions of multi-target estimators can be found in \cite[Sec. 14.5]{mahler_book_2007} and \cite[Sec. 6]{GarciaFernandezWGS:2018}.

For performance evaluation of extended object estimates with ellipsoidal extents, a comparison study has shown that among six compared performance measures, the Gaussian Wassterstein Distance (\gwd) metric is the best choice \cite{YangBG:2016}. The \gwd is defined as \cite{GivensS:1984}
\begin{align}
	d_{\rm GW}\left(\sx,\hat{\sx}\right) = & \| H\bm{\xi} - H\hat{\bm{\xi}} \|^{2} \\
	& + \tr\left(\ext+\hat{\ext}-2 \left( \ext^{\frac{1}{2}} \hat{\ext} \ext^{\frac{1}{2}} \right)^{\frac{1}{2}} \right), \nonumber
\end{align}
where the measurement model $H$ picks out the position from the state vector. The \gwd single target metric is integrated into the Generalised Optimal Sub-Pattern Assignment (\gospa) multi object metric \cite{RahmathullahGFS:2017}, defined as
\begin{align}
	d_{p}^{(c,2)}(\setX,\hat{\setX}) = & \Bigg[ \min_{\theta\in\Theta^{(|\setX|,|\hat{\setX}|)}} \sum_{(i,j)\in\theta} d_{\rm GW}^{(c)}(\sx^{i},\hat{\sx}^{j})^{p} \bigg. \nonumber \\
	& \left. + \frac{c^{p}}{2} \left( |\setX| - |\theta| + |\hat{\setX}|-|\theta| \right) \right]^{\frac{1}{p}}
	\label{eq:GOSPAmetricDefinition}
\end{align}
where $d_{\rm GW}^{(c)}(\sx^{i},\hat{\sx}^{j}) = \min(c\ ; \ d_{\rm GW}(\sx^{i},\hat{\sx}^{j}))$, $\Theta^{(|\setX|,|\hat{\setX}|)}$ is the set of all possible 2D assignment sets, $c$ denotes the cut-off at base distance, and $p$ determines the severity of penalizing the outliers in the localisation component. Here we use $c=10$, $p=1$.

The \gospa metric was proposed in \cite{RahmathullahGFS:2017} as a generalisation of the \ospa metric \cite{SchuhmacherVV:2008} that allows a decomposition of the error into three parts: 1) localisation error $ \sum_{(i,j)\in\theta} d_{\rm GW}(\sx^{i},\hat{\sx}^{j})^{p}$, 2) missed targets $\frac{c^{p}}{2} \left( |\setX| - |\theta| \right)$, and 3) false targets $\frac{c^{p}}{2} ( |\hat{\setX}|-|\theta| )$.
For the simulation study, we present results for the following:
\begin{enumerate}
	\item \gospa \eqref{eq:GOSPAmetricDefinition};
	\item normalised localisation error (\textsc{nle}): localisation error, divided by the number of localised targets, i.e., average localisation error;
	\item number of missed targets (\textsc{mt});
	\item number of false targets (\textsc{ft});
	\item number of cardinality errors (\textsc{ce}): sum of \textsc{mt} and \textsc{ft};
	\item Time: the total time to process one full sequence of measurement sets.
\end{enumerate}
\gospa is important as a single unified performance metric; the other quantities are presented because they represent properties that are important in \mtt.

\begin{figure}
	\centering
	\includegraphics[width=0.32\columnwidth]{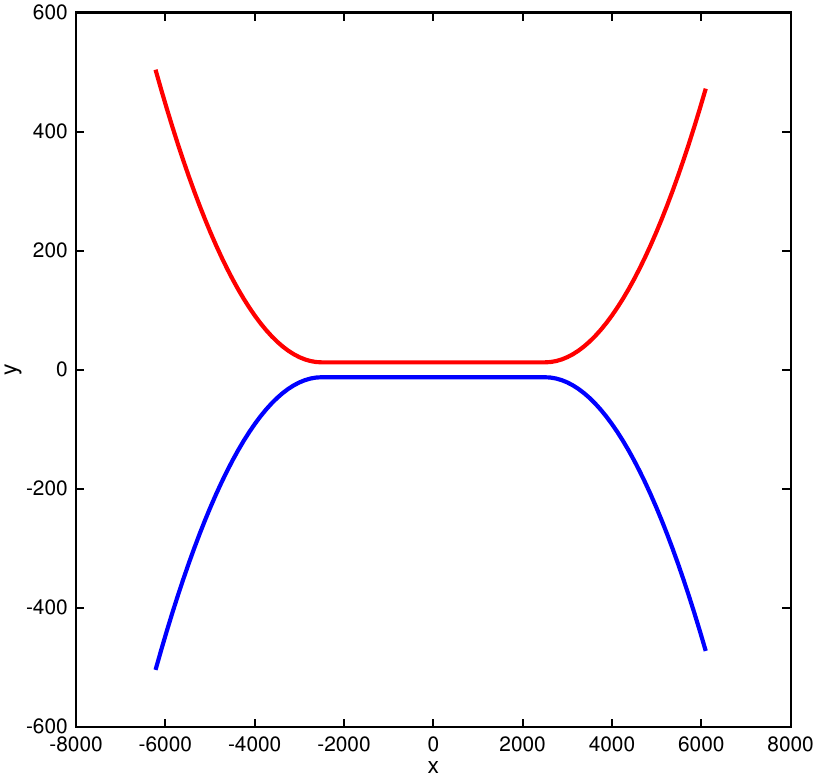}
	\includegraphics[width=0.32\columnwidth]{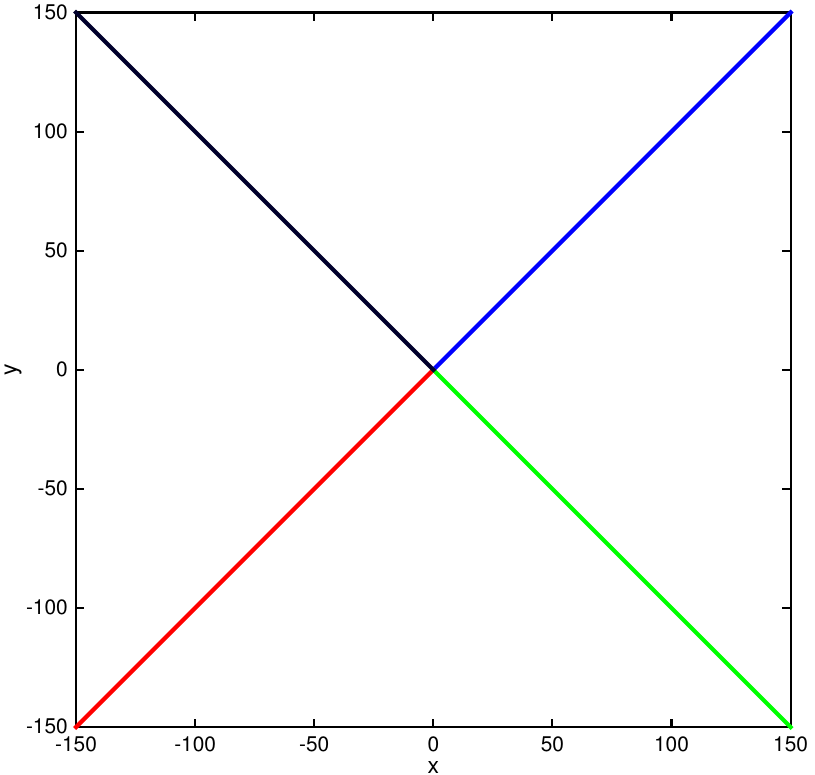}
	\includegraphics[width=0.32\columnwidth]{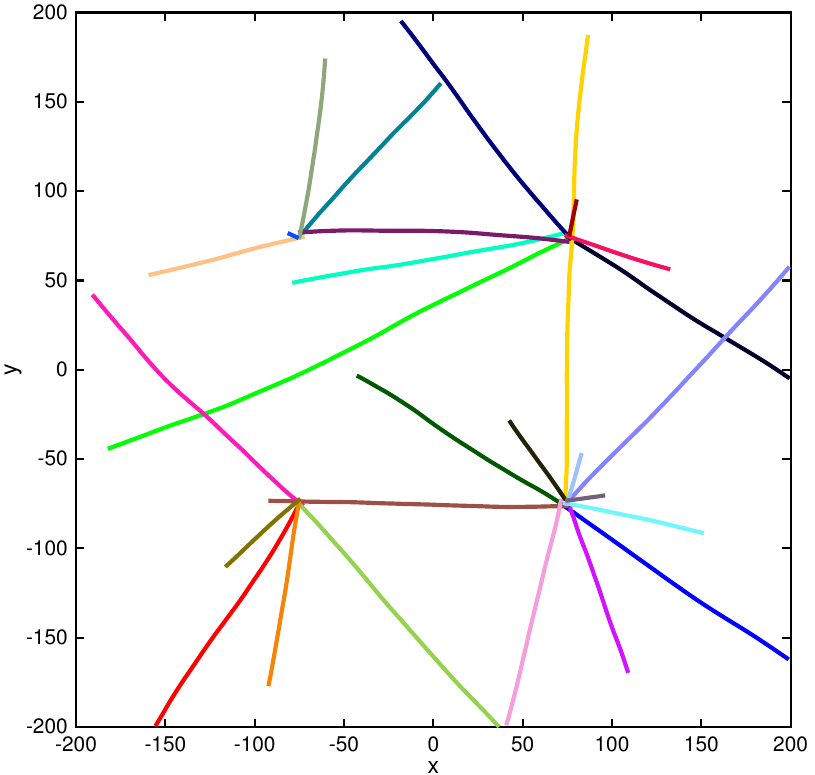}
	\caption{True target tracks for the three simulated scenarios. In scenario 1 (left), the targets are born well separated, move close to each other, and then split. In scenario 2 (center), the targets are born from the same location, but at different times. In scenario 3 (right), there are four different birth locations.}
	\label{fig:TrueTracks}
\end{figure}

\subsection{Simulation study}
Three scenarios were simulated; the first two have been used in previous work to evaluate extended target tracking, see \cite{GranstromLO:2012,GranstromO:2012a,LundquistGO:2013,BeardRGVVS:2016}, the third was constructed for this paper. For each scenario, 100 Monte Carlo runs were performed, and the presented results are averaged over the Monte Carlo runs.

\begin{figure}
	\centering
	\includegraphics[width=1.0\columnwidth]{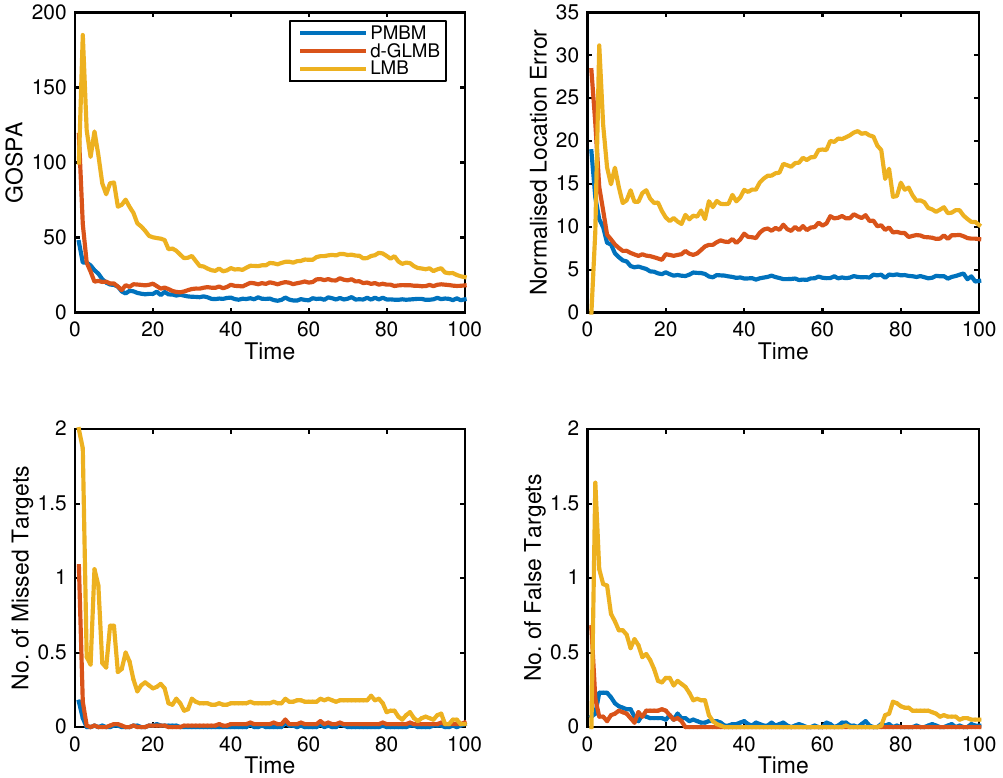}
	\caption{Results for simulation scenario 1.}
	\label{fig:GOSPA_scen1}
\end{figure}

In the first scenario, two targets are simulated for $100$ time steps. The true trajectories are shown in Figure~\ref{fig:TrueTracks}, the true measurement rates were $10$ and $20$. The scenario parameters were set to $p_{\rm D}=0.98$, $p_{\rm S}=0.99$, and $\lambda=30$. This scenario is challenging because when the targets are close their measurements form a single cluster, making the data association difficult. The \gospa performance is shown in Figure~\ref{fig:GOSPA_scen1}, and the numbers are summarised in Table~\ref{tab:scenario_1_results}. Overall, for this scenario the \gospa results show that the \pmbm filter gives smaller errors than both the \dglmb filter and the \lmb filter. Noteworthy is that the localisation error of the \pmbm filter is unaffected when the targets are close and the data association is more complicated, whereas both the \dglmb filter and the \lmb filter show increased localisation errors when the targets are close.

\begin{figure}
	\centering
	\includegraphics[width=1.0\columnwidth]{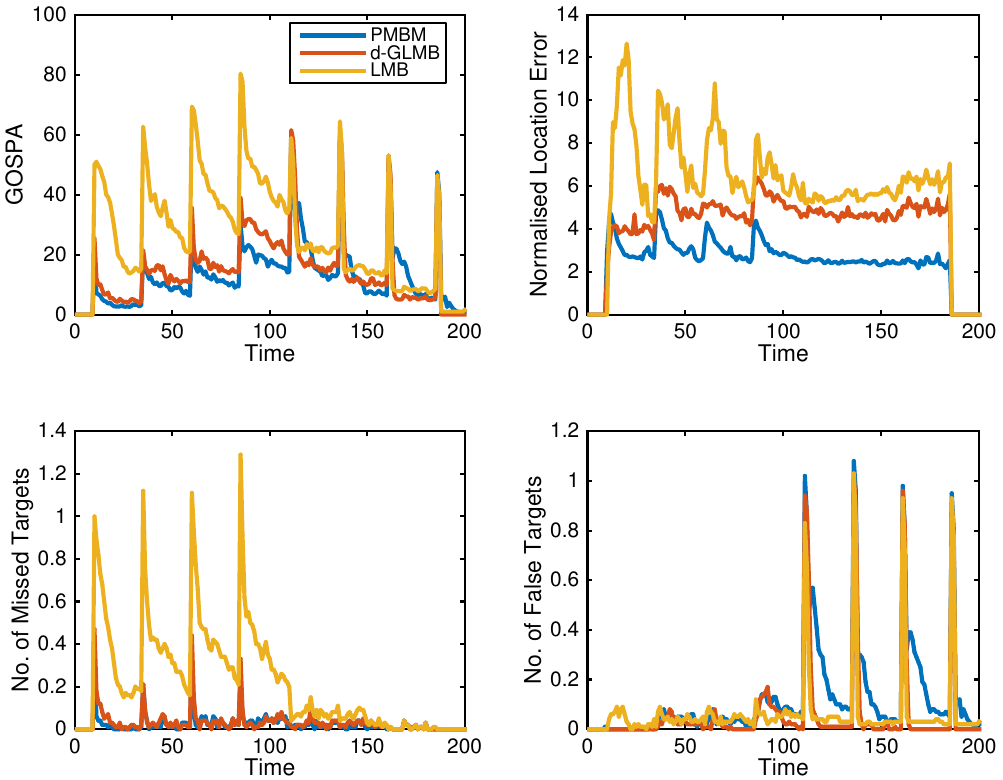}
	\caption{Results for simulation scenario 2.}
	\label{fig:GOSPA_scen3}
\end{figure}

In the second scenario, four targets were simulated for $200$ time steps. The true trajectories are shown in Figure~\ref{fig:TrueTracks}, the true measurement rates were $4$, $6$, $8$, and $10$. The scenario parameters were set to $p_{\rm D}=0.80$, $p_{\rm S}=0.99$, and $\lambda=30$. The targets appear at different times from the same birth location, and disappear at different times. This scenario illustrates how the different filters handle target birth and target death. The \gospa performance is shown in Figure~\ref{fig:GOSPA_scen3}, and the numbers are summarised in Table~\ref{tab:scenario_2_results}. The results show that for most time steps, the \pmbm filter has lower \gospa error. The \pmbm filter has lower \textsc{nle} and lower \textsc{mt}, however, it is also slower at removing targets that have disappeared, which can be seen in the \textsc{ft} results.

In the third scenario, $27$ randomly generated targets were simulated for $100$ time steps. The true trajectories are shown in Figure~\ref{fig:TrueTracks}, the true measurement rates were, for each target, randomly selected from $\{7,8,9\}$.  The targets appear in, and disappear from, the surveillance area at different time steps. The parameters were set to $p_{\rm D}=0.90$, $p_{\rm S}=0.99$, and $\lambda=60$. The birth spatial density consists of four \ggiw components, with positions in $[\pm 75 \ , \ \pm 75]^{\tp}$.  This scenario illustrates how the different filters handle a higher target number and higher clutter density. The \gospa performance is shown in Figure~\ref{fig:GOSPA_scen2}, and the numbers are summarised in Table~\ref{tab:scenario_3_results}. 

In the third scenario, the \lmb filter has larger \gospa error, larger \textsc{nle}, significantly larger \textsc{mt}, and lowest \textsc{ft}. Considering \gospa error, Figure~\ref{fig:GOSPA_scen2} shows that \dglmb is slightly lower than \pmbm until about time 70, when \dglmb starts to increase and becomes larger than \pmbm. The \textsc{nle} and \textsc{mt} are approximately the same for the \pmbm filter and the \dglmb filter, except after time step 80 when there is a quite small difference in favour of the \pmbm filter. For \textsc{ft}, the \pmbm filter gives larger errors until about time step 70, when \textsc{ft} becomes larger for the \dglmb filter. The \pmbm filter's false targets are due to targets that disappear; the \pmbm filter is slightly slower at removing disappeared targets, an effect that could also be observed in the second scenario. The increase in \textsc{ft} for the \dglmb filter starts around time step $60$, when the number of targets in the scene increases to $10$ or more. The rapid increase around time $80$ corresponds to when the number of targets become larger than $15$.

\begin{table}
	\caption{Results for simulation Scenario 1}
	\label{tab:scenario_1_results}
	\centering
	\begin{tabular}{c | c | c | c }
		Filter 		& \pmbm 			& \dglmb 				& \lmb \\
		\hline
		\gospa 		& $\bf 1166$	& $2020$		& $4411$ \\
		\textsc{nle} 	& $\bf 948$	& $1742$		& $2350$ \\
		\textsc{mt}		& $\bf 0.63$	& $2.88$		& $23.88$ \\
		\textsc{ft} 		& $3.73$		& $\bf 2.67$	& $17.34$ \\
		\textsc{ce}	& $\bf 4.36$	& $ 5.55$		& $41.22$ \\
		Time 		& $46$		& $177$		& $\bf 7$ 
	\end{tabular}
\end{table}

The computational cost of the \pmbm filter is significantly lower than the cost of the \dglmb filter, but higher than the cost of the \lmb filter. The \lmb filter is faster than the \pmbm filter because it maintains a single \mb density, as opposed to the \pmbm which has a mixture of \mb densities. However, the single \mb density is also why the \lmb filter has largest \gwd-\gospa error. That the \pmbm is significantly faster than the \dglmb filter is mainly due to two reasons: 1) the \pmbm has uncertain target existence, whereas the \dglmb has certain target existence, and 2) the \pmbm is unlabelled which permits merging of similar \mb densities. 

To conclude, the simulation study shows that for the compared scenarios the \pmbm filter achieves an appealing compromise between the computational cost and the tracking performance.

\begin{table}
	\caption{Results for simulation Scenario 2}
	\label{tab:scenario_2_results}
	\centering
	\begin{tabular}{c | c | c | c }
		Filter 		& \pmbm 				& \dglmb 				& \lmb \\
		\hline
		\gospa 		& $\bf 2574$		& $2865$		& $5274$ \\
		\textsc{nle} 	& $\bf 1149$		& $1963$		& $2381$ \\
		\textsc{mt}		& $\bf 4.54$		& $6.16$		& $44.15$ \\
		\textsc{ft} 		& $23.96$			& $\bf 11.88$	& $13.7$ \\
		\textsc{ce}	& $28.5$			& $\bf 18.04$	& $57.85$ \\
		Time 		& $32$				& $287$		& $\bf 8$ 
	\end{tabular}
\end{table}

\subsection{Experiment}
An experiment with data from a 2D lidar sensor was performed. This data set has previously been used for tracking using the \ggiw-\phd filter \cite{GranstromO:2012a}. The tracking results for detected targets are essentially identical for this data, since the measurements have relatively low measurement noise and there are very few clutter detections. Instead, the challenges posed by this data, and laser range data in general, are caused by occlusions since a Lidar cannot see behind a target. Because of this we emphasize here the estimation of the \ppp density for undetected targets. 

The data, shown in Figure~\ref{fig:Lidar}, contains four pedestrians, two of which remain in the surveillance area for a longer time. One pedestrian moves to the center of the surveillance area and remains there for the remainder of the experiment. The other pedestrian walks around in the surveillance area, both behind and in front of the first pedestrian. A pragmatic approach to handling the occlusions is to use a heterogeneous and time-variant probability of detection $p_{\rm D}(\sx)$. Such an approach was used in \cite{GranstromO:2012a}, and it makes it possible to keep track of targets while they are occluded. The method from \cite{GranstromO:2012a} is used here, and the \ppp intensity for undetected targets correctly captures the increased likelihood that a yet undetected target is located in the occluded part of the surveillance area. 

Results are shown in Figure~\ref{fig:LidarResultsPPP}, where the position component of the undetected \ppp intensity is shown. The area behind the stationary target is occluded for an extended period of time, and the \ppp intensity correctly captures that in this area we can expect there to be undetected targets. The remaining parts of the surveillance volume, which is not occluded, has very low intensity, which is consistent with our expectation that there is not any undetected targets there.

\section{Conclusions}
\label{sec:Conclusions}
This paper has presented a Poisson multi-Bernoulli mixture conjugate prior for tracking of multiple extended targets. Due to the unknown data associations, the complexity of the update is prohibitive, however, standard \mtt methods such as gating and clustering can easily be used to handle this. A \ggiw implementation is also presented, for tracking of extended targets with elliptic shapes. A simulation study shows that the \pmbm filter compares well to the extended target \dglmb and \lmb filters.

An experiment with  laser range data illustrated how the Poisson process helps us to model undetected targets. That is, by approximating the probability of detection, the tracking filter can both track detected targets during occlusions, and represent parts of the surveillance area where yet undetected targets may be located. There are many more scenarios where the probability of detection varies in both time and space, creating a need to model undetected targets. Examples include sensors that scan the surveillance area in a non-deterministic way, such as  radars with narrow lobes that can be focused on certain bearings, or optical sensors mounted on airborne vehicles.

Lastly, note that labels can be used to form target trajectories from the output of the \lmb and \dglmb filters. Using the \pmbm filter output to form target trajectories is a topic for future work.

\appendix

In this appendix we prove Theorem~\ref{thm:Update} using the probability generating transform (pgfl). 

\begin{figure}
	\centering
	\includegraphics[width=1.0\columnwidth]{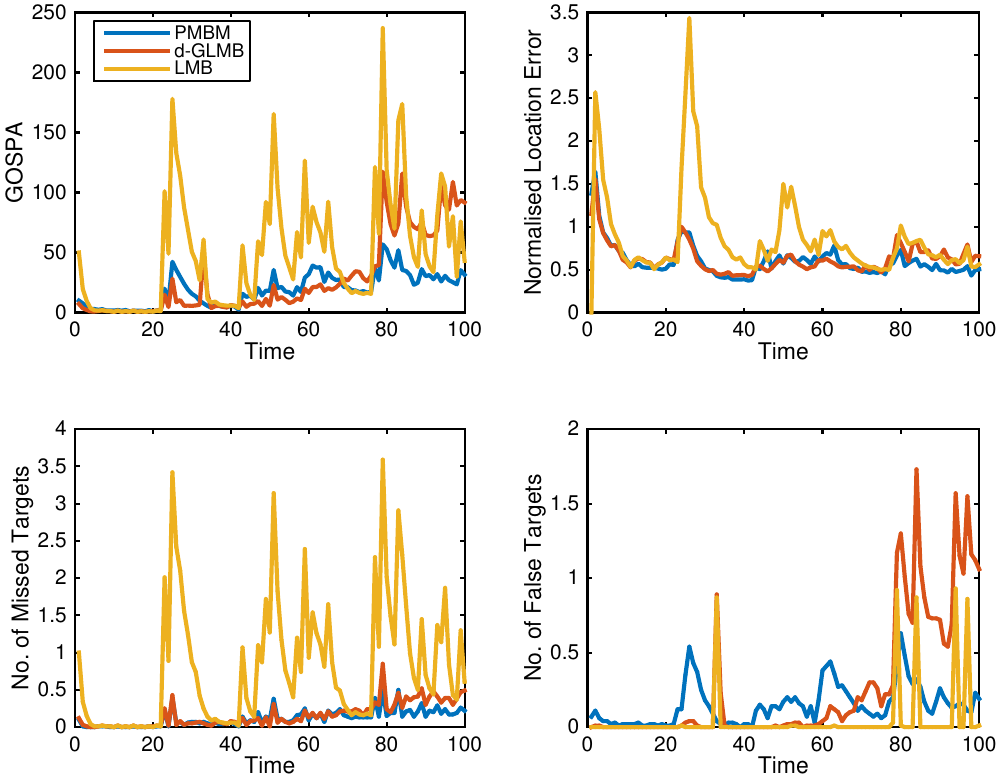}
	\caption{Results for simulation scenario 3.}
	\label{fig:GOSPA_scen2}
\end{figure}

\begin{table}
	\caption{Results for simulation Scenario 3}
	\label{tab:scenario_3_results}
	\centering
	\begin{tabular}{c | c | c | c }
		Filter 		& \pmbm 		& \dglmb 		& \lmb \\
		\hline
		\gospa 		& $\bf 1892$	& $2753 $		& $4919 $ \\
		\textsc{nle} 	& $\bf 556$	& $617 $		& $711 $ \\
		\textsc{mt}		& $\bf 12.06$	& $16.18 $	& $79.62 $ \\
		\textsc{ft} 		& $14.66 $	& $26.55 $	& $\bf 4.54 $ \\
		\textsc{ce}	& $\bf 26.72 $	& $42.73 $	& $84.16 $ \\
		Time 		& $89 $		& $1450 $		& $\bf 11 $ 
	\end{tabular}
\end{table}

\begin{figure}
	\centering
	\includegraphics[width=1.0\columnwidth]{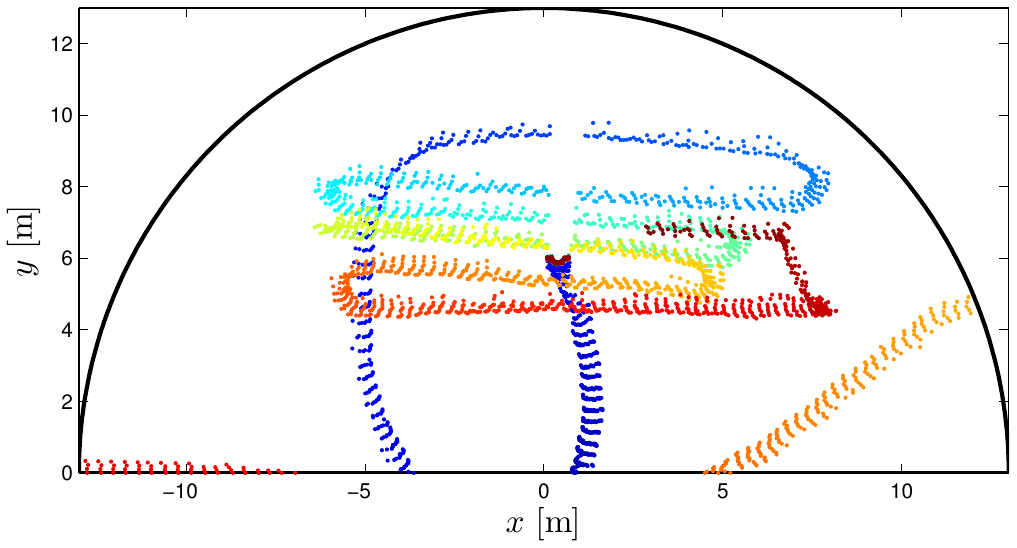}
	\caption{2D laser range data, colors are used to visualise different time steps.}
	\label{fig:Lidar}
\end{figure}

\begin{figure}
	\centering
	\includegraphics[width=0.5\columnwidth]{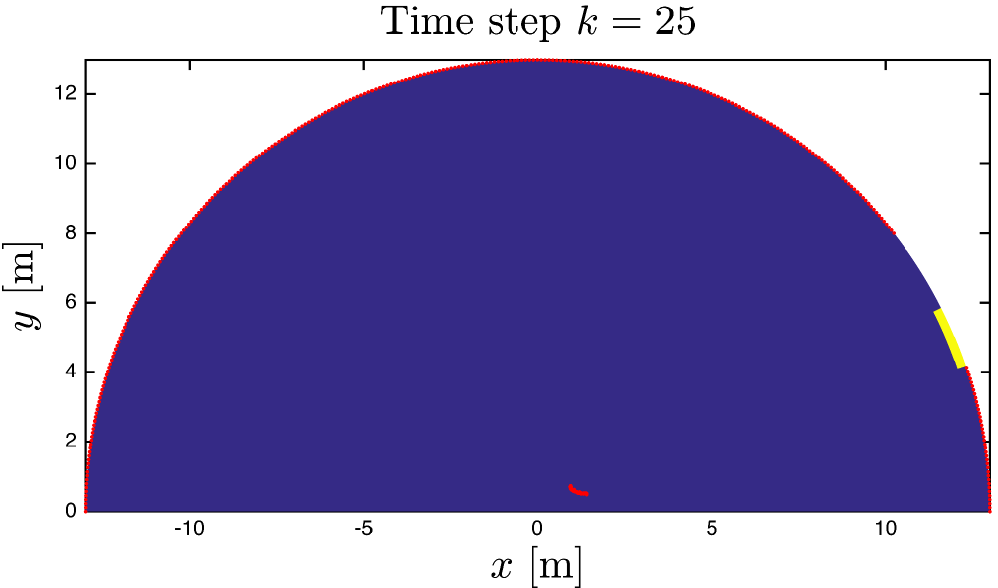}\includegraphics[width=0.5\columnwidth]{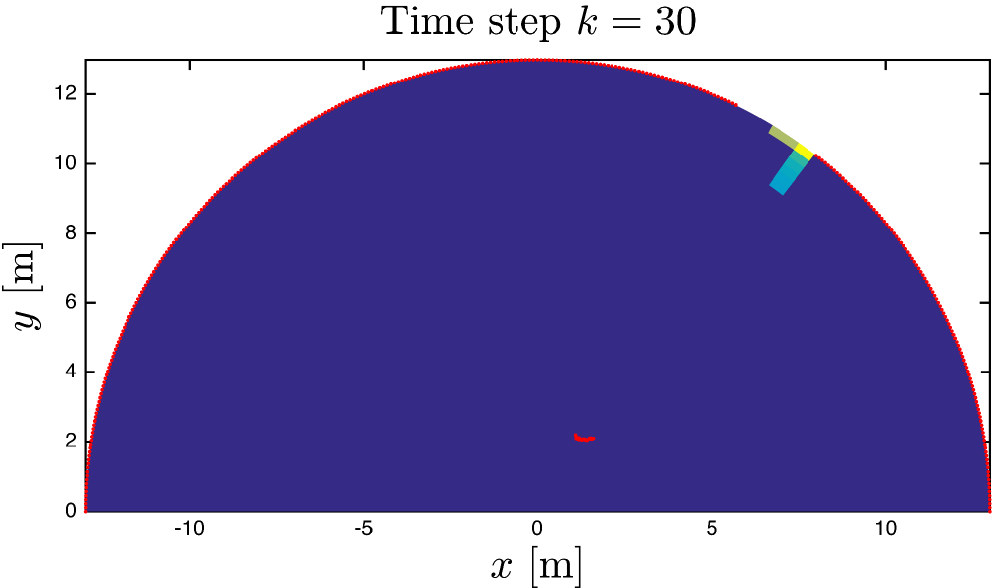}
	\includegraphics[width=0.5\columnwidth]{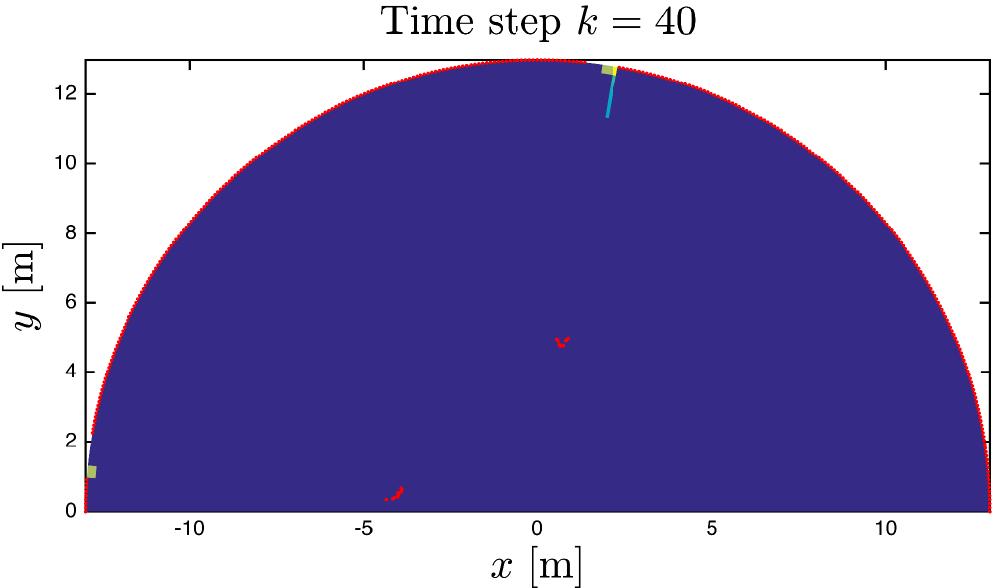}\includegraphics[width=0.5\columnwidth]{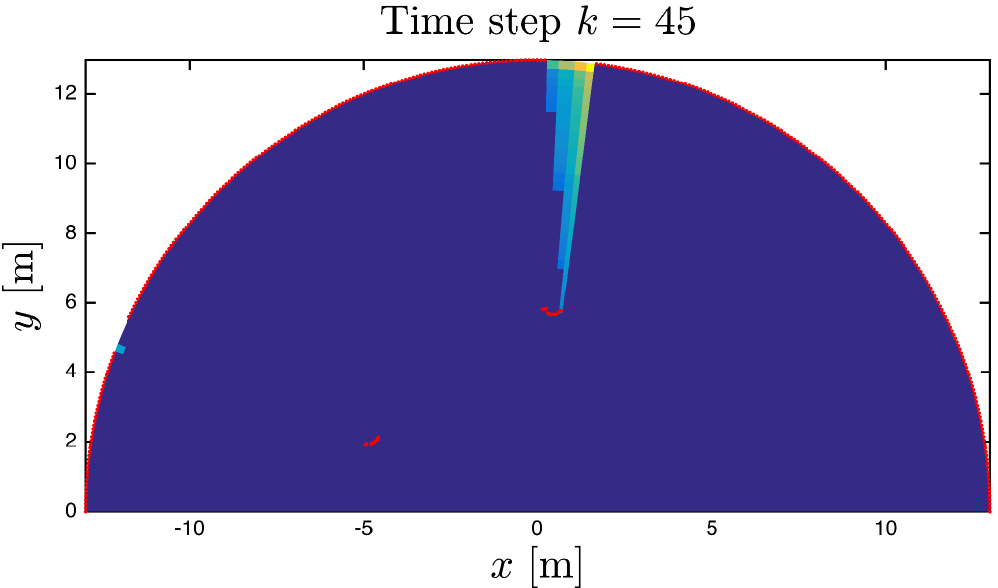}
	\includegraphics[width=0.5\columnwidth]{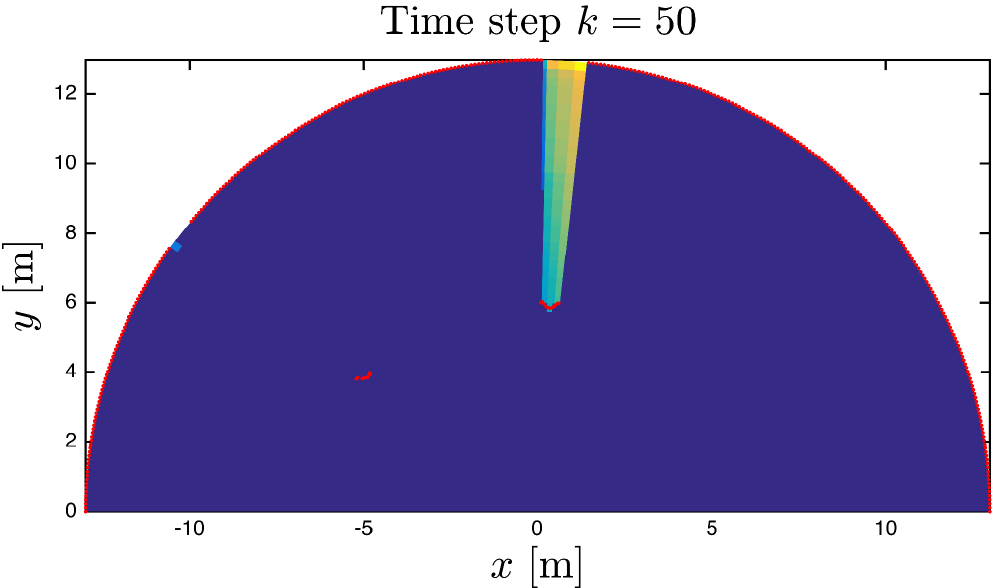}\includegraphics[width=0.5\columnwidth]{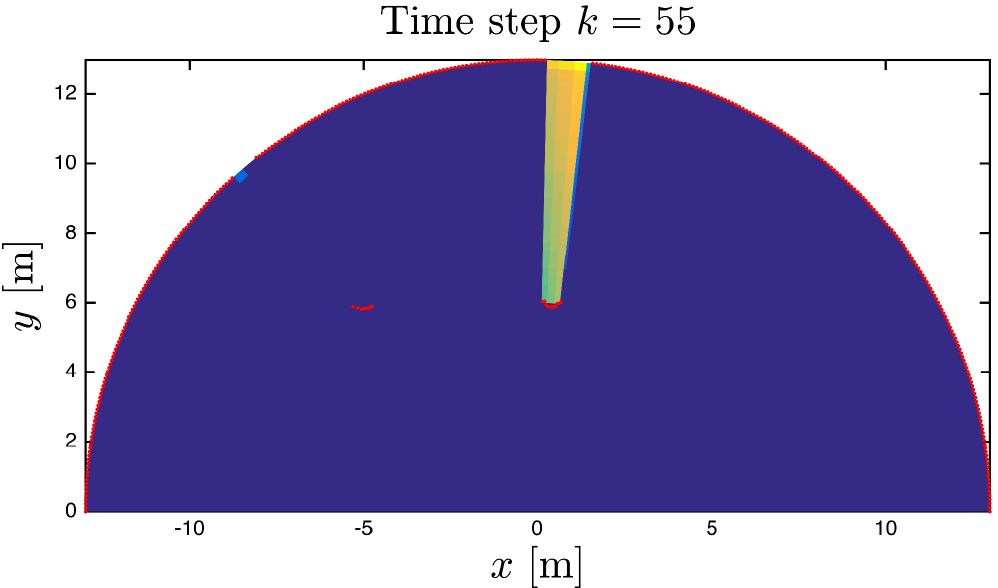}
	\includegraphics[width=0.5\columnwidth]{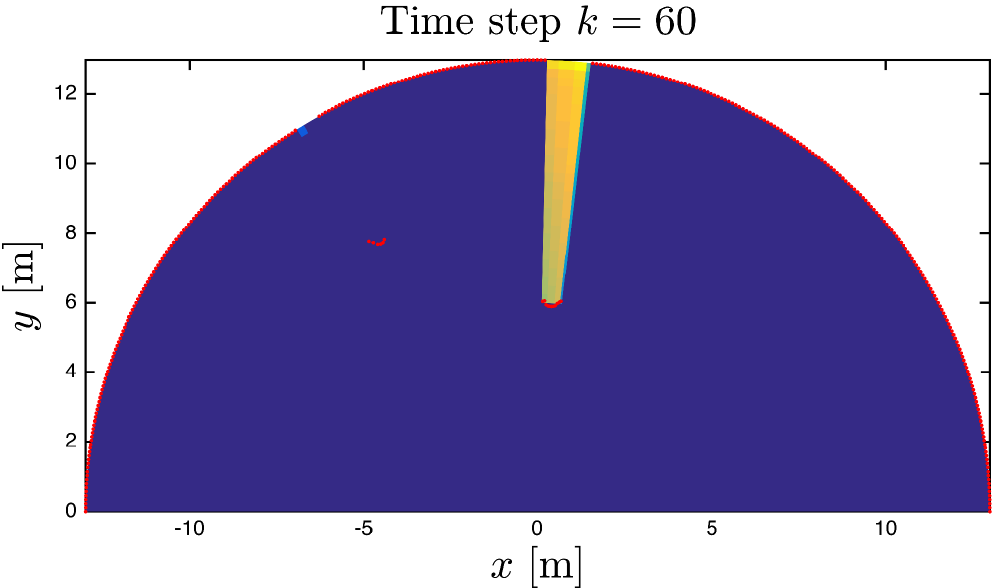}\includegraphics[width=0.5\columnwidth]{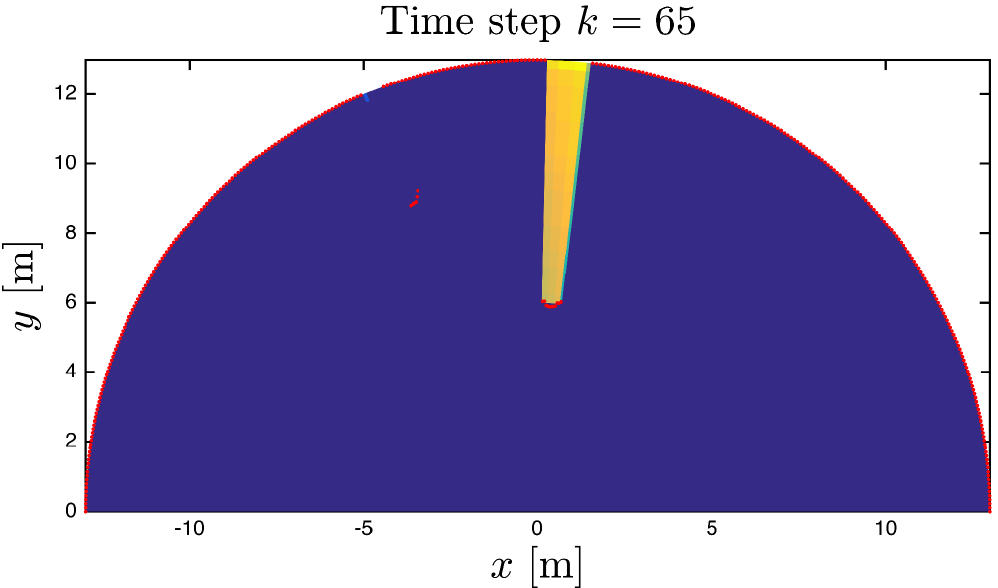}
	\caption{\pmbm tracking results, visualising the position component of the \ppp intensity for undetected targets. The red dots are the lidar detections, blue color corresponds to low intensity, green corresponds to intermediate intensity, and yellow high intensity. When the pedestrian remains stationary, the \ppp intensity increases behind them.}
	\label{fig:LidarResultsPPP}
\end{figure}

\section{Proof of update}
\label{app:UpdateProof}


\subsection{ PGFL background }
In this subsection we give a brief background on the pgfl. Let $\setX$ be an \rfs with multi-object density $f(\setX)$. The probability generating functional (pgfl) is a multitarget integral transform of the multi-object density. The pglf and its inverse are defined as \cite{mahler_book_2007}
\begin{subequations}
\begin{align}
G[h] = & \int h^{\setX} f(\setX)\delta\setX, \label{eq:PGFLtransform} \\
f(\setX) = & \left. \frac{\delta}{\delta \setX}G[h]\right|_{h=0} , \label{eq:PGFLinversetransform} \\
\frac{\delta}{\delta \setX}G[h] = & \frac{\delta^{|\setX|}}{\prod_{\sx\in\setX}\delta\sx}G[h] \label{eq:setDerivative} \\
\frac{\delta}{\delta \sx}G[h] \triangleq  &  \lim_{\varepsilon \searrow 0} \frac{G[h+\varepsilon \delta_{\sx}] - G[h]}{\varepsilon}
\end{align}
\end{subequations}
where $h(\sx)$ is a test-function. The \ppp, Bernoulli, and multi Bernoulli pgfls, corresponding to the densities \eqref{eq:PPPdensity}, \eqref{eq:BernoulliSetDensity}, and \eqref{eq:MultiBernoulliDensity}, respectively, are given by \cite{mahler_book_2007}
\begin{align}
	G^{\rm PPP}[h] & = \exp\left( \conv{D}{h} - \conv{D}{1}\right) , \\
	G^{\rm Ber}[h] & = 1-r + r \conv{f}{h} , \\
	G^{\rm MB}[h]  = &  \prod_{i\in\mathbb{I}} \left( 1- r^{i}+ r^{i} \conv{f^{i}}{h} \right) .
\end{align}
It follows from \eqref{eq:PGFLtransform} that the \mbm pgfl is
\begin{align}
G^{\rm MBM}[h] = & \sum_{j\in\mathbb{J}} \mathcal{W}^{j} \prod_{i\in\mathbb{I}^{j}} \left( 1-r^{j,i} + r^{j,i} \conv{f^{j,i}}{h} \right) .
\end{align}
Due to the standard independence assumption, see Section~\ref{sec:BernoulliRFS}, the pgfl of the \pmbm density \eqref{eq:PMBMsetDensity} is given by
\begin{subequations}%
\begin{align}%
G[h] = & G^{u}[h] \sum_{j\in\mathbb{J}} \mathcal{W}^{j} G^{j}[h] ,\\
G^{u}[h] = & \exp \left( \conv{D^{u}}{h} -  \conv{D^{u}}{1} \right) ,\\
G^{j}[h]  = & \prod_{i\in\mathbb{I}^{j}} \left( 1 - r^{j,i} + r^{j,i} \conv{f^{j,i}}{h}\right) .
\end{align}%
\label{eq:PosteriorPMBMpgfl}%
\end{subequations}%
We rewrite this as
\begin{subequations}%
\begin{align}%
G[h] & = \sum_{j\in\mathbb{J}} \mathcal{W}^{j} G^{u}[h] \prod_{i\in\mathbb{I}^{j}}  G^{j,i}[h] , \\
G^{j,i}[h]  & = 1 - r^{j,i} + r^{j,i} \conv{f^{j,i}}{h} .
\end{align}%
\end{subequations}%

\subsection{PGFL form of Bayes update}
In this subsection we present the Bayes update on pgfl form. Let $G_{+}[h]$ be the prior pgfl that corresponds to the multi-target density $f_{+}(\setX) = f_{k|k-1}(\setX_{k} | \setZ^{k-1})$ in \eqref{eq:MultiobjectPrediction}. The pgfl for the measurement model (Section~\ref{sec:StandardMeasurementModel}) with \ppp clutter and \ppp target measurements is
\begin{subequations}
\begin{align}
G[g|\setX] = & G^{\rm C}[g]  \left(1-p_{\rm D} + p_{\rm D}\exp\left( \conv{\gamma\phi}{g-1} \right) \right)^{\setX} , \label{eq:StandardMeasurementModelPGFL} \\
G^{\rm C}[g] = & \exp\left(\conv{\kappa}{g-1} \right) .
\end{align}%
\end{subequations}
where $g(\sz)$ is a test-function. The joint target and measurement pgfl is
\begin{subequations}
\begin{align}
F&[g,h] = \int h^{\setX} G[g|\setX] f_{+}(\setX ) \delta \setX \\
& = G^{\rm C}[g] G_{+}[h\left(1-p_{\rm D} + p_{\rm D}\exp\left( \gamma \conv{\phi}{g} - \gamma \right) \right)].\label{eq:MeasurementModelPGFLspecific}
\end{align}%
\label{eq:MeasurementModelPGFL}%
\end{subequations}%
Assuming that the prior pgfl $G_{+}[h]$ is \pmbm \eqref{eq:PosteriorPMBMpgfl} , and inserting into the joint pgfl \eqref{eq:MeasurementModelPGFL} gives
 \begin{subequations}
\begin{align}
F[g,h] = & \sum_{j\in\mathbb{J}} \mathcal{W}^{j} F^{j}[g,h] ,\\
F^{j}[g,h] = & F^{{\rm C}u}[g,h] \prod_{i\in\mathbb{I}^{j}} F^{j,i}[g,h] ,\\
F^{{\rm C}u}[g,h] = & G^{\rm C}[g] G_{+}^{u}[h (1-p_{\rm D} + p_{\rm D} e^{ \gamma \conv{\phi}{g} - \gamma } )] ,\\
F^{j,i}[g,h] = & G_{+}^{j,i}[h (1-p_{\rm D} + p_{\rm D} e^{ \gamma \conv{\phi}{g} - \gamma } )] .
\end{align}%
\label{eq:JointPGFLthreeParts}%
\end{subequations}%
The updated pgfl $G[h]$ that corresponds to the Bayes updated density $f_{k|k}(\setX_k|\setZ^k) $ in \eqref{eq:MultiobjectCorrection} is given by \cite[pp. 530--531, s. 14.8.2]{mahler_book_2007}
\begin{align}
G[h] = & \frac{\left. \frac{\delta F[g,h]}{\delta \setZ} \right|_{g=0} }{ \left. \frac{\delta F[g',h']}{\delta \setZ} \right|_{g'=0,h'=1}  } . \label{eq:UpdatedPGFLdifferentiation}
\end{align}


\subsection{Preliminaries to proof of Theorem~\ref{thm:Update}}

In this subsection, we establish some preliminary results that will allow us to obtain the differentiation $\frac{\delta F[g,h]}{\delta \setZ}$ that is needed in \eqref{eq:UpdatedPGFLdifferentiation}. From the sum rule for functional derivatives \cite[p. 395]{mahler_book_2007}, it holds that $\frac{\delta F[g,h]}{\delta \setZ} = \sum_{j\in\mathbb{J}} \mathcal{W}^{j} \frac{\delta F^{j}[g,h]}{\delta \setZ}$. From the product rule for functional derivatives \cite[p. 395]{mahler_book_2007}, it follows that the differentiation $\frac{\delta F^{j}[g,h]}{\delta \setZ}$ consists of combinations of differentiations of $F^{{\rm C}u}[g,h]$ and of $F^{j,i}[g,h]$.

\begin{lemma}\label{lem:ExpDiff}\it
	\begin{align}
		\frac{\delta \conv{s}{\exp(\gamma\conv{\phi}{g})} }{\delta \sz} = \conv{s}{\gamma\phi_{\sz}\exp(\gamma\conv{\phi}{g})}
	\end{align} 
	where $s(\sx)$ is any function of $\sx$ and $\phi_{\sz} = \phi(\sz | \sx)$.
\end{lemma}
Successive application of the chain rule for functional derivatives \cite[p. 395]{mahler_book_2007} gives that
\begin{subequations}
\begin{align}
	\frac{\delta \conv{s}{\exp(\gamma\conv{\phi}{g})} }{\delta \sz}  & = \conv{s}{ \frac{\delta \exp(\gamma\conv{\phi}{g})}{\delta \sz} } \\
	& = \conv{s}{ \gamma \phi_{\sz} \exp(\gamma\conv{\phi}{g}) }.
\end{align}%
\end{subequations}%
This concludes the proof of Lemma~\ref{lem:ExpDiff}.

\begin{lemma}\label{lem:PPPpgflDiff} \it The differentiation of $F^{{\rm C}u}[g,h]$ w.r.t. a single measurement $\sz$ is
\begin{align}
	\frac{\delta F^{{\rm C}u}[g,h]}{\delta \sz}
	& = \left(\kappa(\sz) + \conv{h}{ D_{+}^{u} \ell_{\sz} e^{\gamma\conv{\phi}{g}}}\right) F^{{\rm C}u}[g,h] , \label{eq:PPPpgflDiffAux1}
\end{align}
and the differentiation of $\kappa(\sz) + \conv{h}{ D_{+}^{u} \ell_{\sz} e^{\gamma\conv{\phi}{g}}}$ w.r.t. a set of measurements $\setY$ is
\begin{align}
	\frac{\delta \left(\kappa(\sz) + \conv{h}{ D_{+}^{u} \ell_{\sz} e^{\gamma\conv{\phi}{g}}}\right)}{\delta \setY}  = \conv{h}{ D_{+}^{u} \ell_{\sz\cup\setY} e^{\gamma\conv{\phi}{g}}}. \label{eq:PPPpgflDiffAux2}
\end{align}
\end{lemma}
The proof of \eqref{eq:PPPpgflDiffAux1} is given in \cite[eq. 31-32]{mahler_FUSION_2009_extTarg}. For the proof of \eqref{eq:PPPpgflDiffAux2}, it follows from the sum rule for functional derivatives \cite[p. 395]{mahler_book_2007} that
\begin{align}
	\frac{\delta \left(\kappa(\sz) + \conv{h}{ D_{+}^{u} \ell_{\sz} e^{\gamma\conv{\phi}{g}}}\right)}{\delta \setY} = \frac{\delta \conv{h}{ D_{+}^{u} \ell_{\sz} e^{\gamma\conv{\phi}{g}}}}{\delta \setY}. \label{eq:PPPpgflDiffAux3}
\end{align}
and the equivalence of the \textsc{rhs} of \eqref{eq:PPPpgflDiffAux3} and the \textsc{rhs} of \eqref{eq:PPPpgflDiffAux2} follows from Lemma~\ref{lem:ExpDiff} and the definition of set derivative \eqref{eq:setDerivative}. This concludes the proof of Lemma~\ref{lem:PPPpgflDiff}

\begin{lemma}\label{lem:BERpgflDiff} \it The differentiation of $F^{j,i}[g,h]$ w.r.t. a measurement set $\setZ$ is
\begin{align}
	\frac{\delta F^{j,i}[g,h]}{\delta \setZ} & =  r_{+}^{j,i}\conv{h}{f_{+}^{j,i} \ell_{\setZ} e^{\gamma\conv{\phi}{g}}}.
\end{align}
\end{lemma}
It follows from the defintion of $\conv{\cdot}{\cdot}$, see Table~\ref{tab:notation}, and the sum rule for functional derivatives \cite[p. 395]{mahler_book_2007} that
\begin{align}
	\frac{\delta F^{j,i}[g,h]}{\delta \setZ}
	& = r_{+}^{j,i}  \frac{\delta \left(\conv{f_{+}^{j,i} h p_{\rm D} e^{-\gamma }  }{ e^{ \gamma \conv{\phi}{g}} } \right)}{\delta \setZ}
\end{align}
This concludes the proof of Lemma~\ref{lem:BERpgflDiff}.

\begin{lemma}\label{lem:PMBpgflDiff} \it The differentiation of $F^{j}[g,h]$ w.r.t. $\setZ$ is
\begin{subequations}
\begin{align}
	\frac{\delta F^{j}[g,h]}{\delta \setZ} & = F^{{\rm C}u}[g,h] \sum_{\assoc \in \assocspace^{j}}  \prod_{C \in \assoc} F^{'}_{C}[g,h]
\end{align}
where
\begin{align}
	& F^{'}_{C}[g,h] =  \\
	& \left\{\begin{array}{ccc} \text{\footnotesize $\kappa^{\altcell_{C}} + \conv{h}{ D_{+}^{u} \ell_{\altcell_{C}} e^{\gamma\conv{\phi}{g}}}$} & \text{\footnotesize if} & \text{\footnotesize $C\cap\mathbb{I}_{+}^{j}=\emptyset, |\altcell_{C}|=1$} \\ \text{\footnotesize $\conv{h}{ D_{+}^{u} \ell_{\altcell_{C}} e^{\gamma\conv{\phi}{g}}}$} & \text{\footnotesize if} & \text{\footnotesize $C\cap\mathbb{I}_{+}^{j}=\emptyset, |\altcell_{C}|>1$} \\ \text{\footnotesize $F^{j,i_{C}}[g,h]$} & \text{\footnotesize if} & \text{\footnotesize $C\cap\mathbb{I}_{+}^{j} \neq \emptyset, \altcell_{C}=\emptyset$} \\ \text{\footnotesize $r_{+}^{j,i_{C}} \conv{h}{f_{+}^{j,i_{C}} \ell_{\altcell_{C}} e^{\gamma \conv{\phi}{g}} }$} & \text{\footnotesize if} & \text{\footnotesize $C\cap\mathbb{I}_{+}^{j}\neq\emptyset,\altcell_{C}\neq\emptyset$} .\end{array} \right. \nonumber 
\end{align}%
\label{eq:ProofDifferentiationEq}%
\end{subequations}
\end{lemma}
The proof is by induction: For the initial step, assume that $\mathbb{M}=\{m_1\}$. Differentiation, Lemma~\ref{lem:PPPpgflDiff} and Lemma~\ref{lem:BERpgflDiff}, give
\begin{align}
	\frac{\delta F^{j}[g,h]}{\delta \sz^{m_1}}  = &   \label{eq:PMBpgflDiff} \\
	 F^{{\rm C}u}[g,h] & \bigg[ \left(\kappa(\sz) + \conv{h}{D_{+}^{u} \ell_{\sz} e^{\gamma\conv{\phi}{g}}}\right) \prod_{i\in\mathbb{I}_{+}^{j}}F^{j,i}[g,h]  \nonumber \bigg. \\
	& \bigg. + \sum_{\hat{\imath} \in \mathbb{I}_{+}^{j}}  r_{+}^{j,\hat{\imath}}\conv{h}{f_{+}^{j,\hat{\imath}} \ell_{\setZ} e^{\gamma\conv{\phi}{g}}} \prod_{i\in\mathbb{I}_{+}^{j},i\neq\hat{\imath}}F^{j,i}[g,h] \bigg]. \nonumber
\end{align}
We see that \eqref{eq:PMBpgflDiff} is consistent with \eqref{eq:ProofDifferentiationEq}: we have the partitions of $\{m_1\} \cup \mathbb{I}_{+}^{j}$, for which there is at most one $i\in\mathbb{I}_{+}^{j}$ in each cell. The first row corresponds to a partition in which $m_1$ is placed in a cell of its own, $\{m_1\},\{i_1\},\ldots,\{i_I\}$, i.e., an association in which none of the previously detected targets are detected, and the single measurement is either from clutter or a new target. The second row corresponds to partitions $\{i_1\},\ldots,\{m_1,\hat{\imath}\},\ldots,\{i_I\}$, i.e., associations where the single measurement is associated to one of the existing targets.

Now, assume that we have established \eqref{eq:ProofDifferentiationEq} for 
\begin{align}
	\mathbb{M}_{-} = \{m_1,\ldots,m_{M-1}\}
\end{align}
and that we are to establish \eqref{eq:ProofDifferentiationEq} for
\begin{align}
	\mathbb{M} = \{m_1,\ldots,m_{M-1},m_{M}\} .
\end{align}
For the sake of notational clarity, let $\assocspace^{j}_{-}$ be the association space corresponding to the index set $\mathbb{M}_{-}$, and let $\assocspace^{j}$ be the association space corresponding to the index set $\mathbb{M}$. Differentiation gives
\begin{align}
	& \frac{\delta F^{{\rm C},u}[g,h] \sum_{\assoc \in \assocspace^{j}_{-}} \prod_{C \in \assoc} F^{'}_{C}[g,h]}{\delta \sz^{m_{M}}} \label{eq:LemmaGeneralInductionStep} \\
	= &  F^{{\rm C}u}[g,h] \bigg[  \sum_{\assoc \in \assocspace^{j}_{-}}  \sum_{\hat{C}\in\assoc} \frac{\delta F^{'}_{\hat{C}}[g,h]}{\delta \sz^{m_{M}}} \prod_{C \in \assoc,C\neq\hat{C}} F^{'}_{C}[g,h] \bigg. \nonumber \\
	& \bigg. + \left(\kappa(\sz^{m_{M}}) + \conv{h}{D_{+}^{u} \ell_{\sz^{m_{M}}} e^{\gamma\conv{\phi}{g}}}\right)   \sum_{\assoc \in \assocspace^{j}_{-}}\prod_{C \in \assoc} F^{'}_{C}[g,h]  \bigg] . \nonumber
\end{align}
The first row corresponds to new association partitions formed by adding the measurement index $m_{M}$ to one of the existing cells in an association $\assoc\in\assocspace^{j}_{-}$, and the second row corresponds to new association partitions formed by putting $m_{M}$ into a new cell and adding this cell to $\assoc\in\assocspace_{-}^{j}$. Together, this constitutes a summation over all possible ways to partition $\mathbb{M}\cup\mathbb{I}^{j}_{+}$, i.e., a summation over all associations in the association space $\assocspace^{j}$, and  \eqref{eq:LemmaGeneralInductionStep} is thus consistent with  \eqref{eq:ProofDifferentiationEq}. This concludes the proof of Lemma~\ref{lem:PMBpgflDiff}.

\begin{lemma}\label{lem:BernoulliParams} \it  For scalar $a$, scalar $b$, function $c$ and test function $h$, the following relation holds
\begin{subequations}
\begin{align}
	a + b\conv{h}{c} = \mathcal{L} \left( 1 - r + r\conv{h}{f}\right),
\end{align}
where
\begin{align}
	\mathcal{L}  = a + b\conv{c}{1}, \quad r = \frac{b\conv{c}{1}}{a + b \conv{c}{1}}, \quad f = \frac{c}{\conv{c}{1}}.
\end{align}
\end{subequations}
\end{lemma}
The proof is trivial.
 

\subsection{Proof of Theorem~\ref{thm:Update}}
In this subsection, we show that a prior \pmbm pgfl $G_{+}[h]$ of the form \eqref{eq:PosteriorPMBMpgfl} and the measurement model \eqref{eq:StandardMeasurementModelPGFL} result in a \pmbm pgfl that corresponds to the \pmbm density given in Theorem~\ref{thm:Update}. Let the set of measurements $\setZ$ be indexed by the index set $\mathbb{M}$, $\setZ=\{\sz^{m}\}_{m\in\mathbb{M}}$. Differentiating, using Lemma~\ref{lem:PMBpgflDiff}, and setting $g=0$, we get
\begin{align}
	\left.\frac{\delta F[g,h]}{\delta \setZ}\right|_{g=0} & = F^{{\rm C},u}[0,h]\sum_{j\in\mathbb{J}} \mathcal{W}^{j} \sum_{\assoc \in \assocspace^{j}}  \prod_{C \in \assoc} F^{'}_{C}[0,h].
\end{align}
where
\begin{align}
	& F^{'}_{C}[0,h] =  \\
	& \left\{\begin{array}{ccc} \text{\footnotesize $\kappa^{\altcell_{C}} + \conv{h}{ D_{+}^{u} \ell_{\altcell_{C}}}$} & \text{\footnotesize if} & \text{\footnotesize $C\cap\mathbb{I}_{+}^{j}=\emptyset, |\altcell_{C}|=1$} \\ \text{\footnotesize $\conv{h}{ D_{+}^{u} \ell_{\altcell_{C}}}$} & \text{\footnotesize if} & \text{\footnotesize $C\cap\mathbb{I}_{+}^{j}=\emptyset, |\altcell_{C}|>1$} \\ \text{\footnotesize $1 -r_{+}^{j,i_{C}} +r_{+}^{j,i_{C}}\conv{h}{ f_{+}^{j,i_{C}} q_{\rm D}}$} & \text{\footnotesize if} & \text{\footnotesize $C\cap\mathbb{I}_{+}^{j} \neq \emptyset, \altcell_{C}=\emptyset$} \\ \text{\footnotesize $r_{+}^{j,i_{C}} \conv{h}{f_{+}^{j,i_{C}} \ell_{\altcell_{C}}}$} & \text{\footnotesize if} & \text{\footnotesize $C\cap\mathbb{I}_{+}^{j}\neq\emptyset,\altcell_{C}\neq\emptyset$} .\end{array}  \right. \nonumber 
\end{align}%
Applying Lemma~\ref{lem:BernoulliParams}, we get
\begin{align}
	& \left.\frac{\delta F[g,h]}{\delta \setZ}\right|_{g=0} \label{eq:PGFLdiffAux1} \\
	& = F^{{\rm C},u}[0,h]\sum_{j\in\mathbb{J}} \sum_{\assoc \in \assocspace^{j}} \mathcal{W}^{j} \prod_{C\in\assoc} \mathcal{L}_{C} \left(1-r_{C} + r_{C}\conv{h}{f_{C}} \right) \nonumber
\end{align}
where
\begin{subequations}
\begin{align}
	& \mathcal{L}_{C} =  \left\{\begin{array}{cl} \text{\footnotesize $\kappa^{\altcell_{C}} + \conv{D_{+}^{u}}{  \ell_{\altcell_{C}}}$} & \text{\footnotesize if $C\cap\mathbb{I}_{+}^{j}=\emptyset, |\altcell_{C}|=1$} \\ \text{\footnotesize $\conv{D_{+}^{u}}{ \ell_{\altcell_{C}}}$} & \text{\footnotesize if $C\cap\mathbb{I}_{+}^{j}=\emptyset, |\altcell_{C}|>1$} \\ \text{\footnotesize $1 -r_{+}^{j,i_{C}} +r_{+}^{j,i_{C}}\conv{ f_{+}^{j,i_{C}} }{ q_{\rm D}}$} & \text{\footnotesize if $C\cap\mathbb{I}_{+}^{j} \neq \emptyset, \altcell_{C}=\emptyset$} \\ \text{\footnotesize $r_{+}^{j,i_{C}} \conv{f_{+}^{j,i_{C}}}{ \ell_{\altcell_{C}}}$} & \text{\footnotesize if $C\cap\mathbb{I}_{+}^{j}\neq\emptyset,\altcell_{C}\neq\emptyset$} \end{array} \right. \\
	& r_{C} = \left\{\begin{array}{cl} \text{\footnotesize $\frac{\conv{D_{+}^{u}}{  \ell_{\altcell_{C}}}}{\kappa^{\altcell_{C}} + \conv{D_{+}^{u}}{  \ell_{\altcell_{C}}}}$} & \text{\footnotesize if $C\cap\mathbb{I}_{+}^{j}=\emptyset, |\altcell_{C}|=1$} \\ \text{\footnotesize $1$} & \text{\footnotesize if $C\cap\mathbb{I}_{+}^{j}=\emptyset, |\altcell_{C}|>1$} \\ \text{\footnotesize $\frac{r_{+}^{j,i_{C}}\conv{ f_{+}^{j,i_{C}} }{ q_{\rm D}}}{1 -r_{+}^{j,i_{C}} +r_{+}^{j,i_{C}}\conv{ f_{+}^{j,i_{C}} }{ q_{\rm D}}}$} & \text{\footnotesize if $C\cap\mathbb{I}_{+}^{j} \neq \emptyset, \altcell_{C}=\emptyset$} \\ \text{\footnotesize $1$} & \text{\footnotesize if $C\cap\mathbb{I}_{+}^{j}\neq\emptyset,\altcell_{C}\neq\emptyset$} \end{array} \right.  \\
	& f_{C}(\sx) = \left\{\begin{array}{cl} \text{\footnotesize $\frac{D_{+}^{u}(\sx)\ell_{\altcell_{C}}(\sx)}{\conv{ D_{+}^{u}}{ \ell_{\altcell_{C}}}}$} & \text{\footnotesize if $C\cap\mathbb{I}_{+}^{j}=\emptyset$} \\ \text{\footnotesize $\frac{f_{+}^{j,i_{C}}(\sx) q_{\rm D}(\sx)}{\conv{f_{+}^{j,i_{C}}}{q_{\rm D}}}$} & \text{\footnotesize if $C\cap\mathbb{I}_{+}^{j} \neq \emptyset, \altcell_{C}=\emptyset$} \\ \text{\footnotesize $\frac{f_{+}^{j,i_{C}}(\sx) \ell_{\altcell_{C}}(\sx)}{\conv{f_{+}^{j,i_{C}} }{ \ell_{\altcell_{C}}}}$} & \text{\footnotesize if $C\cap\mathbb{I}_{+}^{j}\neq\emptyset,\altcell_{C}\neq\emptyset$} .\end{array}  \right. 
\end{align}%
\label{eq:UpdatedMBMparametersProof}%
\end{subequations}
Setting $h=1$ we get
\begin{align}
	& \left.\frac{\delta F[g,h]}{\delta \setZ}\right|_{g=0,h=1} = F^{{\rm C},u}[0,1]\sum_{j\in\mathbb{J}} \sum_{\assoc \in \assocspace^{j}} \mathcal{W}^{j} \prod_{C\in\assoc} \mathcal{L}_{C} \label{eq:PGFLdiffAux2}
\end{align}
and taking the ratio of \eqref{eq:PGFLdiffAux1} and \eqref{eq:PGFLdiffAux2}, cf. \eqref{eq:UpdatedPGFLdifferentiation}, we get the pgfl of the Bayes updated density,
\begin{align}
	G_{}&[h] = \frac{F^{{\rm C},u}[0,h]}{F^{{\rm C},u}[0,1]} \label{eq:UpdatedPGFL} \\
	& \times \frac{\sum_{j\in\mathbb{J}} \sum_{\assoc \in \assocspace^{j}} \mathcal{W}^{j} \prod_{C\in\assoc} \mathcal{L}_{C} \left(1-r_{C} + r_{C}\conv{h}{f_{C}} \right) }{\sum_{j\in\mathbb{J}} \sum_{\assoc \in \assocspace^{j}} \mathcal{W}^{j} \prod_{C\in\assoc} \mathcal{L}_{C}} \nonumber 
\end{align}
where the ratio
\begin{subequations}
\begin{align}
\frac{F^{{\rm C}u}[0,h]}{F^{{\rm C}u}[0,1]} & = \exp\left\{ \conv{D_{}^{u}}{h} - \conv{D_{}^{u}}{1} \right\} \label{eq:updatedPPPpgfl}  \\
D_{}^{u}(\sx) & = q_{\rm D}(\sx)D_{+}^{u}(\sx). \label{eq:updatedPPPintensityProof}
\end{align}
\end{subequations}
We see that $G[h]$ in \eqref{eq:UpdatedPGFL} is a product of \eqref{eq:updatedPPPpgfl}, which is the pgfl of a \ppp with intensity \eqref{eq:updatedPPPintensityProof}, and the pgfl of a \mbm with parameters \eqref{eq:UpdatedMBMparametersProof}. This is consistent with Theorem~\ref{thm:Update}, and concludes the proof.




\bibliographystyle{IEEEtran}
\bibliography{PHD_ext_targ_track}
%
%
%

\end{document}